\documentclass[reprint,prl,twocolumn,showpacs,superscriptaddress,aps,longbibliography]{revtex4-1}
\usepackage[utf8]{inputenc}
\usepackage{graphicx}
\usepackage[colorlinks=true,citecolor=blue]{hyperref}
\usepackage{amsmath}
\usepackage{amssymb}
\usepackage{soul}
\usepackage{bm}
\usepackage{gensymb}
\usepackage{amsfonts,amssymb,amsbsy}
\usepackage{siunitx}
\usepackage{float}
\usepackage{physics}
\usepackage{booktabs}
\usepackage[percent]{overpic}

\newcommand{\dIdV}{d$I$/d$V$~}
\usepackage{xcolor}
\usepackage{natbib}

\usepackage{lineno}

\begin{document}

\title{Emergent ferromagnetism in the NiI$_2$-NbSe$_2$ van der Waals heterostructure}

\author{Büşra Gamze Arslan}
\affiliation{Aalto University, Department of Applied Physics, 00076 Aalto, Finland}

\author{Mohammad Amini}
\affiliation{Aalto University, Department of Applied Physics, 00076 Aalto, Finland}

\author{Ziying Wang}
\affiliation{Aalto University, Department of Applied Physics, 00076 Aalto, Finland}

\author{Alessandro Orsini}
\affiliation{Aalto University, Department of Applied Physics, 00076 Aalto, Finland}

\author{Aleš Cahlík}
\affiliation{Aalto University, Department of Applied Physics, 00076 Aalto, Finland}

\author{Jose L. Lado}
\affiliation{Aalto University, Department of Applied Physics, 00076 Aalto, Finland}

\author{Adolfo O. Fumega}
\email{Corresponding authors. Email: adolfo.oterofumega@aalto.fi, robert.drost@aalto.fi, peter.liljeroth@aalto.fi}
\affiliation{Aalto University, Department of Applied Physics, 00076 Aalto, Finland}

\author{Robert Drost}
\email{Corresponding authors. Email: adolfo.oterofumega@aalto.fi, robert.drost@aalto.fi, peter.liljeroth@aalto.fi}
\affiliation{Aalto University, Department of Applied Physics, 00076 Aalto, Finland}

\author{Peter Liljeroth}
\email{Corresponding authors. Email: adolfo.oterofumega@aalto.fi, robert.drost@aalto.fi, peter.liljeroth@aalto.fi}
\affiliation{Aalto University, Department of Applied Physics, 00076 Aalto, Finland}

\begin{abstract}

Multiferroicity arising from non-collinear spin textures and strong spin–orbit interactions offers a route to magnetoelectric functionality in the monolayer limit. Although theory predicts that the properties of monolayer multiferroics can be tuned by strain, gating, or proximity effects, experimental demonstrations of such control remain scarce. Here we show that the magnetic ground state of monolayer NiI$_2$, a prototypical two-dimensional multiferroic, is altered by proximity to a superconducting NbSe$_2$ substrate. Using low-temperature scanning tunnelling microscopy (STM) and spectroscopy (STS), we show that the metallic substrate renormalizes the exchange interactions within NiI$_2$ and drives it into a ferromagnetic ground state. This can be visualized by probing the Yu–Shiba–Rusinov (YSR) states within the superconducting gap of the NbSe$_2$ substrate. Our results establish YSR states as an in situ probe of two-dimensional magnetism and demonstrate substrate engineering as a means of controlling magnetic order in atomically thin materials.

\end{abstract}

\date{\today}

\maketitle 
\section*{Introduction}

The field of magnetic two-dimensional materials was initiated by the discovery of intrinsic ferromagnetism in Cr$_2$Ge$_2$Te$_6$ and CrI$_3$ \cite{Gong2017DiscoveryCrystals,Huang2017Layer-dependentLimit}, demonstrating that long-range magnetic order can persist down to the monolayer limit. Subsequent work rapidly expanded the range of known 2D magnetic phases \cite{Fei2018Two-dimensionalFe3GeTe2,Bedoya-Pinto2021IntrinsicMonolayer,Jiang2018ControllingDoping,Burch2018MagnetismMaterials,Gibertini2019MagneticHeterostructures}, including candidate magnetic quantum materials \cite{Grubisic-Cabo2025RoadmapMaterials}. A more recent addition to this family is the class of 2D helimagnets, where non-collinear spin textures combined with spin–orbit interactions generate an electric polarization, giving rise to multiferroic order in two dimensions \cite{Tokura2014MultiferroicsOrigin,Ju2021Possiblesub2/sub,Song2022EvidenceMultiferroic,Fumega2022,Amini2024AtomicScalesub2/sub, 2601.20713,Miao2025,Wang2026}. Although theoretical studies predict that the properties of these systems could be tuned through strain, electrostatic gating, or the formation of heterostructures, experimental demonstrations of such tunability remain limited.

This tunability could be exploited in van der Waals heterostructures, which combine atomically thin materials with considerable flexibility. Heterostructures with magnetic materials have been used to engineer a variety of exotic quantum phases \cite{Kezilebieke2020TopologicalHeterostructure,Vano2021ArtificialHeterostructure,Wan2023EvidenceLattice, Zhong2017VanValleytronics, Seyler2018ValleyHeterostructures, Song2021DirectMagnets, Chen2023ManipulationHeterostructures}. While these approaches have been widely explored for conventional 2D magnets, similar efforts involving monolayer multiferroics remain largely absent. One challenge is that heterostructure formation can itself modify the magnetic ground state. Exchange interactions in atomically thin magnets are highly sensitive to substrate effects, including charge transfer, hybridization, and electronic screening from the substrate, providing flexible tuning knobs to drive magnetic phase transitions.

Here, we investigate this interplay by growing a two-dimensional multiferroic, NiI$_2$, on a superconducting NbSe$_2$ substrate and probing the resulting heterostructure using low-temperature scanning tunneling microscopy (STM) and spectroscopy (STS). The superconducting substrate provides a sensitive probe of magnetism through Yu–Shiba–Rusinov (YSR) states \cite{Heinrich2018SingleSuperconductors,LoConte2024Magnet-superconductorSuperconductivity}. The exchange interaction between the magnetic moments in NiI$_2$ and the superconductor produces
bound states within the superconducting energy gap, leading to
YSR bands in extended one- or two-dimensional systems \cite{Kezilebieke2020TopologicalHeterostructure,Li2024ObservationHeterostructure,Cuperus2025OneNbSe2,Cuperus2025Non-topologicalHeterostructures, Soldini2023Two-dimensionalSuperconductivity}. Our measurements reveal the absence of the original multiferroic order of single layer NiI$_2$ and instead show YSR bands in the interior of the film together with local density of states (LDOS) signatures at the island edges that are consistent with a ferromagnetic state. These observations can be explained by substrate-induced renormalization
of magnetic interactions, driving a helimagnetic-to-ferromagnetic transition in the NiI$_2$ layer. More broadly, our results demonstrate how YSR states can serve as an in-situ probe of magnetism and highlight substrate engineering as a route for tuning magnetic order in atomically thin systems, potentially enabling controlled manipulation of multiferroic states in future hybrid quantum-material platforms.

\section*{Results}

\subsection*{Growth and properties of NiI$_2$ on NbSe$_2$} 

\begin{figure*}[ht!]
    \centering
    \includegraphics[width=\textwidth]{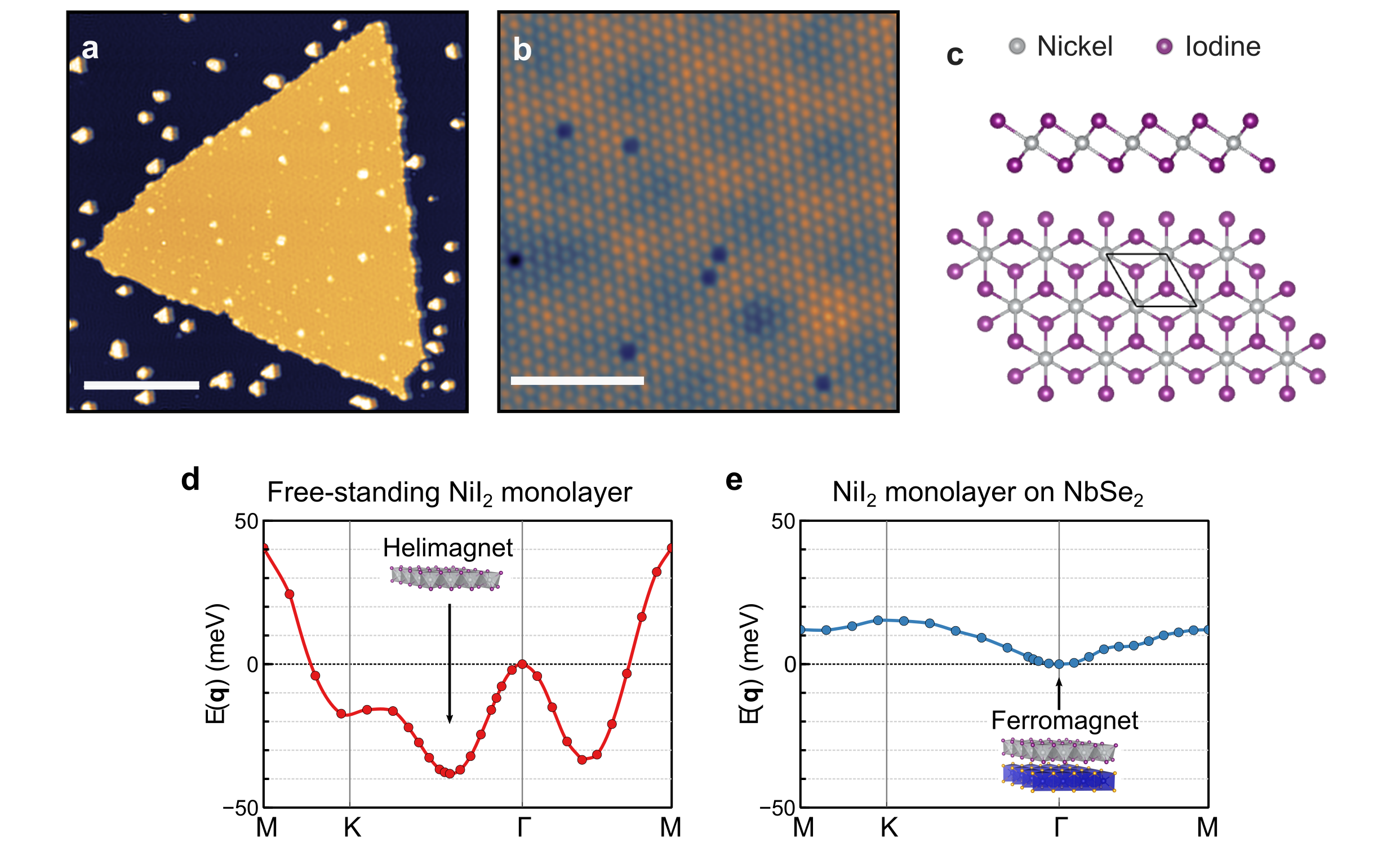}
    \caption{Monolayer NiI$_2$ grown by MBE on NbSe$_2$. \textbf{a,} Large area topography image of a NiI$_2$ island (scale bar 50 nm, setpoint \SI{1}{\volt}, \SI{5.1}{\pico\ampere}). \textbf{b,} Atomic resolution topography image of NiI$_2$ (scale bar 4 nm, setpoint \SI{100}{\milli\volt}, \SI{308}{\pico\ampere}). \textbf{c,} The schematic illustration of monolayer NiI$_2$ with a side and top view. The black rhombohedron represents the unit cell of NiI$_2$. \textbf{d, e,} Energy diagrams of free standing NiI$_2$ and NiI$_2$ on NbSe$_2$ substrate, respectively. Free standing NiI$_2$ shows a helimagnetic ground state while NiI$_2$ on NbSe$_2$ is ferromagnetic.}
    \label{figure1}  
\end{figure*}

We used molecular beam epitaxy (MBE) in ultra-high vacuum (UHV) to grow NiI$_2$ on NbSe$_2$ substrate (see Methods for details). Fig.~\ref{figure1}a shows a typical NiI$_2$ island grown with this method. The islands display a moiré pattern due to the lattice mismatch between NiI$_2$ and NbSe$_2$. Fig.~\ref{figure1}b shows the atomic lattice of NiI$_2$ with a lattice constant of 3.91\,\AA, consistent with the previous studies \cite{Amini2024AtomicScalesub2/sub,Wang2024Orientation-selectiveNiI2,Miao2025,Wang2026}. We also observe several kinds of point defects, most prominently single atom sized dips, which we interpret as iodine vacancies. Fig.~\ref{figure1}c shows and atomic model of NiI$_2$ from side (top) and top (bottom) view where nickel atoms (gray) are sandwiched between iodine atoms (purple), and the black lines mark the unit cell of NiI$_2$.  In contrast to earlier work on a graphite substrates,\cite{Amini2024AtomicScalesub2/sub,Miao2025,Wang2026} we do not observe stripe-like modulation in NiI$_2$ at any bias voltage. On graphite, the stripe pattern arises from the ferroelectric polarization in NiI$_2$ generated by the spin-spiral ground state together with spin-orbit interactions. The absence of ferroelectric stripes indicates that monolayer NiI$_2$ on NbSe$_2$ does not host a non-collinear magnetic ground state, likely stemming from a magnetic phase transition driven by the NbSe$_2$ substrate.

\begin{figure*}[ht!]
    \centering
     \includegraphics[width=\textwidth]{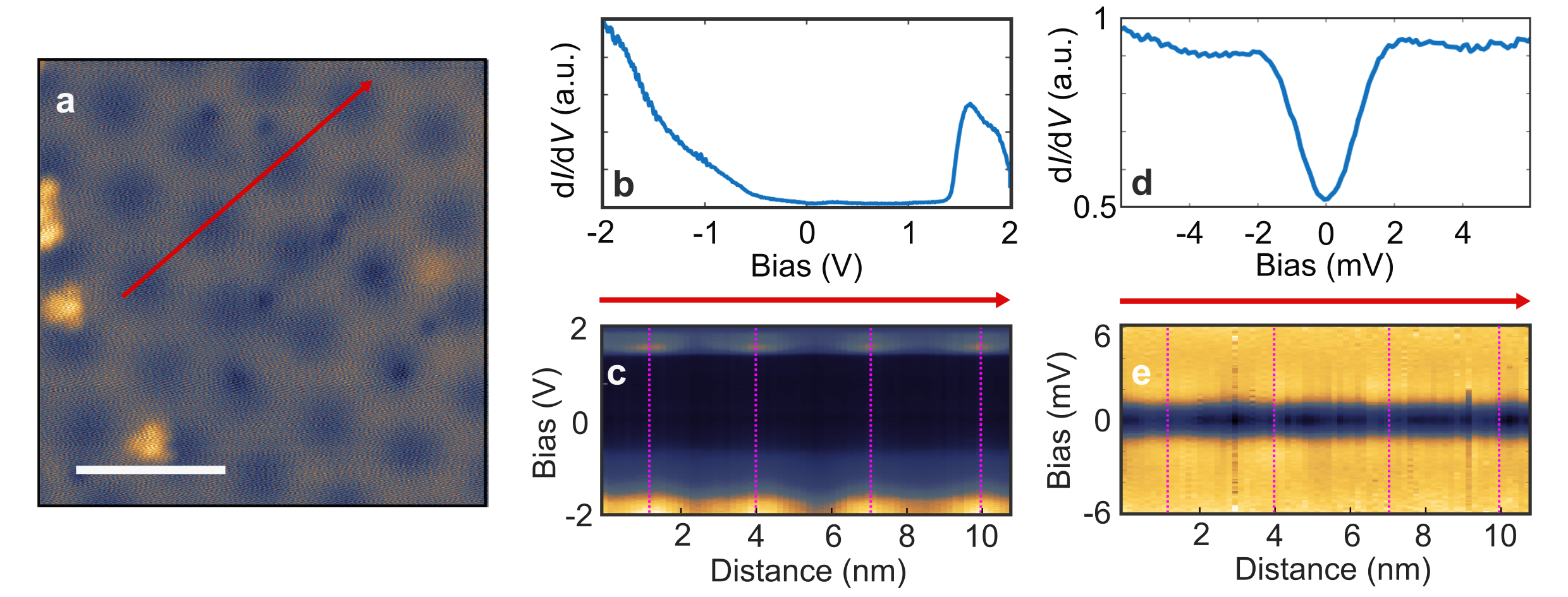}
    \caption{Electronic characterization of NiI$_2$ on NbSe$_2$ at 5\,K. \textbf{a,} Topography image of NiI$_2$ on NbSe$_2$ (scale bar 5 nm, setpoint \SI{1}{\volt}, \SI{315}{\pico\ampere}). \textbf{b,} Typical conductance spectrum of NiI$_2$ on NbSe$_2$ between \SI{2}{\volt} and \SI{-2}{\volt} (stabilization parameters \SI{2}{\volt}, \SI{500}{\pico\ampere}). \textbf{c,} Differential conductance spectra along the line specified with the red arrow in panel a. \textbf{d,} Typical conductance spectrum of the SC gap NbSe$_2$ under NiI$_2$ (stabilization parameters \SI{10}{\milli\volt}, \SI{500}{\pico\ampere}). \textbf{e,} Differential conductance spectra along the red line indicated in panel a (\SI{10}{\milli\volt}, \SI{500}{\pico\ampere}). The magenta dotted lines in panels \textbf{c} and \textbf{e} mark the moiré period, with a slight shift on the last moiré period in \textbf{e} due to the drift.}
    \label{figure2}  
\end{figure*}

While isolated monolayer NiI$_2$ is a spin spiral magnet, proximity to other 2D materials could drive it to a ferromagnetic state. The spin spiral state in NiI$_2$ stems from the competition between ferromagnetic first and antiferromagnetic third nearest neighbor exchange couplings \cite{Fumega2022}, and on e.g.~a graphite substrate, its ground state is close to a ferromagnetic transition. NbSe$_2$ is known to be close to a ferromagnetic instability, and in particular has a dominant ferromagnetic spin susceptibility\cite{PhysRevX.10.041003}. This implies that in the NiI$_2$/NbSe$_2$ heterostructure, NbSe$_2$ could drive a magnetic transition directly from proximity effect, pushing NiI$_2$ towards the ferromagnetic ground state. To demonstrate this, we have performed density functional theory (DFT) calculations on a free-standing NiI$_2$ monolayer and on the NiI$_2$/NbSe$_2$ heterostructure. We use the generalized Bloch theorem to study the energy dependence of the ground state as a function of the spin spiral ground state with magnetization constrained to be $\mathbf m (\mathbf r) = m_0(\cos(\mathbf q \cdot \mathbf r),\sin(\mathbf q \cdot \mathbf r),0)$, where $\mathbf{q}$ is the propagation vector of the spin spiral and $\mathbf r$ are the positions of the Ni sites. For each DFT calculation, $\mathbf q$ is fixed, and $m_0$ is computed self-consistently through the DFT energy minimization. These q-dependent DFT calculations enable us to define the q-dependent ground state $E(\mathbf q)$. The function $E(\mathbf q)$ allows to directly extract the $\mathbf q$ of the real ground state, as the $\mathbf q$ for which $E(\mathbf q)$ is minimal. For a spin-spiral ground state, such as the free-standing NiI$_2$ monolayer (Fig.~\ref{figure1}d), the minimum energy $E(\mathbf q_0)$ appears at finite $ \mathbf q = \mathbf q_0$, where $\mathbf q_0 \approx 2/5\overline{\Gamma K}$ is the physical spin spiral. In contrast, for a ferromagnetic ground state, the minimum appears at $\mathbf q_0 = (0,0)$, and as a result, $E(\mathbf q)$ directly reflects the emergence of a ferromagnetic ground state for monolayer NiI$_2$ on NbSe$_2$ (Fig.~\ref{figure1}e). The emergence of ferromagnetism in NiI$_2$ driven by NbSe$_2$ can be rationalized from a renormalization of the exchange couplings between Ni sites driven by the ferromagnetic RKKY interaction in NbSe$_2$\cite{PhysRevX.10.041003}.

Electronic properties of NiI$_2$ on NbSe$_2$ are experimentally characterized in Fig.~\ref{figure2}, showing the STS data measured at $T=5$~K. Fig.~\ref{figure2}a shows topography image on the island with moiré pattern due to the lattice mismatch between NiI$_2$ and NbSe$_2$. Fig.~\ref{figure2}b shows a d$I$/d$V$ spectrum over a large bias range. Here, the conduction band is observed as a peak around 1.5\,V, while on the HOPG substrate, NiI$_2$ shows the onset of the conduction band at 0.4\,V \cite{Amini2024AtomicScalesub2/sub}. The lattice mismatch between the substrate and the NiI$_2$ layer results in a moiré pattern that periodically modulates electronic properties \cite{Schulz2014EpitaxialTemplate, Joshi2012BoronMonolayer, J.R.Costa2025NanoscaleGraphite,Kezilebieke2022}. As expected, the d$I$/d$V$ spectra acquired along the red arrow in Fig.~\ref{figure2}a show a periodic modulation of the NiI$_2$ conduction band, as illustrated in Fig.~\ref{figure2}c.

Investigating the bulk of NiI$_2$ on NbSe$_2$, we also acquired spatially resolved differential‑conductance measurements across the moiré pattern with the sample bias set in the vicinity of the NbSe$_2$ superconducting gap, shown in Fig.~\ref{figure2}d and \ref{figure2}e. In the waterfall plot of the line spectra shown in Fig.~\ref{figure2}e, a modulation is evident in the appearance of the superconducting gap. The dashed magenta lines in Fig.~\ref{figure2}c, which correspond to modulation of conduction band, coincide with the positions of the reduced apparent gap width observed in the Fig.~\ref{figure2}e. Although it is challenging to resolve in detail at 5\,K, this correlated behavior suggests that the bulk of the island hosts Yu–Shiba–Rusinov (YSR) states, which are likewise modulated by the moiré superlattice.

\subsection*{YSR states in the NiI$_2$ / NbSe$_2$ heterostructure}

\begin{figure*}[ht!]
    \centering
    \includegraphics[width=\textwidth]{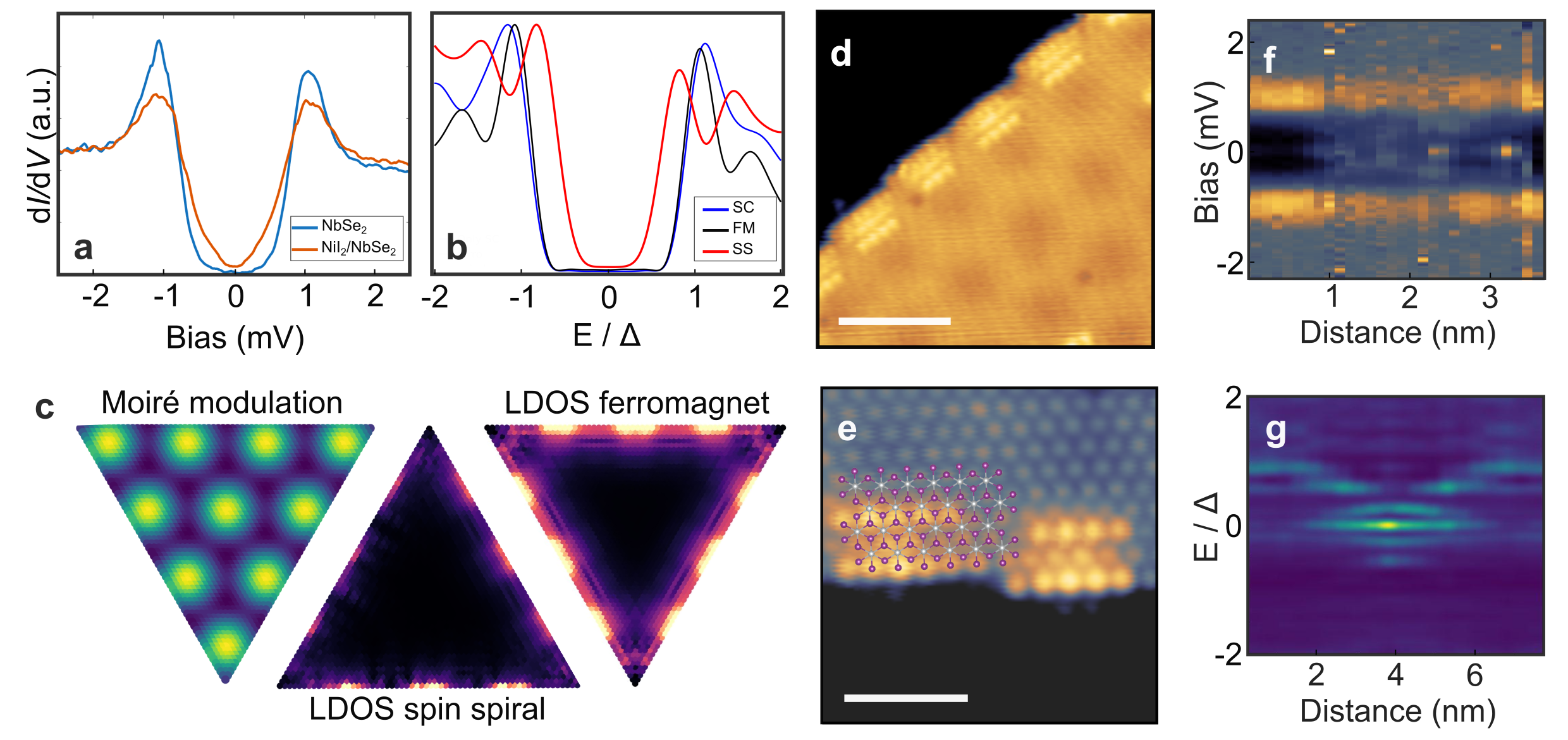}
    \caption{\textbf{a,} Differential conductance spectra measured on NbSe$_2$ substrate (blue curve) and on the NiI$_2$ island (orange curve) at \SI{350}{\milli\kelvin} (stabilization parameters \SI{3}{\milli\volt}, \SI{106.5}{\pico\ampere} and \SI{101}{\pico\ampere}, respectively, modulation voltage $V_m=$ {\SI{50}{\micro\volt}}). \textbf{b,} Calculated LDOS states for ferromagnetic (FM, black curve) and spin spiral (SS, red curve) magnetic ground states. Blue curve shows the calculated conductance spectrum on bare NbSe$_2$ (SC). \textbf{c,} Illustration of the moiré modulation used for the theoretical results, and the calculated LDOS maps for spin spiral and ferromagnetic ground states for a triangular NiI$_2$ island. \textbf{d,} STM image showing the edge contrast and its registry with the moiré pattern  (scale bar 4 nm, $T=5$~K, setpoint \SI{10}{\milli\volt}, \SI{620}{\pico\ampere}). \textbf{e,} Atomic resolution image at the NiI$_2$ island edge and comparison with atomic model of NiI$_2$. Aligning the atoms on topography image with iodine atoms reveals the mismatch between the lattice and LDOS oscillations (scale bar, 2 nm, $T=5$~K, setpoint \SI{10}{\milli\volt}, \SI{603}{\pico\ampere}). \textbf{f,} Differential conductivity measurement acquired along the moiré period at the island edge (\SI{3}{\milli\volt}, \SI{200}{\pico\ampere},  $V_m=$ {\SI{35}{\micro\volt}}) \textbf{g,} The corresponding calculated differential conductivity spectra over the moiré period.}
    \label{figure3}  
\end{figure*}

To probe the behaviour of NiI$_2$ in more detail, we carried out spectroscopy experiments at $T=350$~mK. Fig.~\ref{figure3}a shows differential conductance spectra acquired in 2 different regions. The blue curve shows a reference spectrum acquired on pristine NbSe$_2$, several nm  away from the NiI$_2$ island. This spectrum shows a well-defined superconducting gap, as expected. The orange curve shows a spectrum acquired above a NiI$_2$ island. The spectrum displays a reduced intensity of the coherence peaks with respect to the bare NbSe$_2$ and the gap appears narrower. The gap is filled with new states, which we interpret as a YSR band arising from interaction of magnetic moments in NiI$_2$ with the NbSe$_2$ substrate \cite{Kezilebieke2020TopologicalHeterostructure}. Depending on the exact location of the \dIdV spectrum, the YSR band appear more or less prominently, but the reduced coherence peak intensity is robustly observed. These observations supports the emergence of a ferromagnetic ground state in NiI$_2$ on the NbSe$_2$ substrate.

The idea that the YSR states or bands serve as sensitive probes of the magnetic ground state of NiI$_2$ can be explored more broadly by simulations. We have simulated the expected YSR bands arising in this heterostructure for both a ferromagnetic and spin-spiral magnetic ground state\cite{pyqula}. The 
electronic structure of the 
heterostructure is captured with a low energy model of the form

\begin{equation}
H = H_0 + H_J + H_{SC}
\end{equation}

where 
\begin{equation}
H_0 = \sum_{ij,s}
t_{ij} c^\dagger_{i,s} c_{j,s}
\end{equation}
is the single particle hopping of NbSe$_2$,
\begin{equation}
H_J = \sum_{i,s,s'} \mathbf{J}(\mathbf r_i)
\cdot \bm \sigma^{s,s'} c^\dagger_{i,s} c_{i,s'}
\end{equation}
is the exchange induced coupling of NiI$_2$ in NbSe$_2$, where $\bm J(\mathbf r_i)$
denotes the spatial distribution of the exchange coupling. The term
\begin{equation}
H_{SC} = \Delta \sum_{i} 
c^\dagger_{i,\uparrow} c^\dagger_{i,\downarrow} + \text{h.c.}
\end{equation}
accounts for the intrinsic s-wave superconducting of NbSe$_2$,
leading to a Bogoliubov-de-Gennes Hamiltonian.

The different potential magnetic states of NiI$_2$ can be directly encoded in the functional
form of $\bm J(\mathbf r)$ by taking a parametrization of the form
$
\bm J(\mathbf r) = (\mathcal {J} (\mathbf r) \cos (\mathbf q \cdot \mathbf r),
\mathcal {J}  \sin (\mathbf q \cdot \mathbf r),0)
$, where $\mathbf q$ parametrizes the spin spiral and $\mathcal {J}$ denote the
modulation of the exchange coupling stemming from the moir\'e pattern and
possible enhanced coupling at the island edges (details in the Supplementary Materials, SM). The local density of states can be obtained from the electron
spectral function of the BdG Hamiltonian
defined as $A(\omega,\mathbf r) = \sum_s \langle \bm r,s | \delta (\omega - H) |\bm r,s \rangle$, where $| \bm r,s \rangle\equiv c^\dagger_{\bm r,s}|\Omega\rangle$, with $|\Omega\rangle$
the empty state.

The calculated LDOS spectra and maps can be compared with the experimental spectra and images close to the island edges (Fig.~\ref{figure3}). The different spin configurations have significantly different energy spectra as illustrated in Fig.~\ref{figure3}b (the effect of the various model parameters is explored in more detail in the SM). Calculated spectra with a spin-spiral ground state consistently result in strongly shifted and split coherence peak features in the LDOS. The results corresponding to a ferromagnetic ground state are more in-line with the experimental results (Fig.~\ref{figure3}a). 

The ferromagnetic and spin-spiral ground states are predicted to result in distinct LDOS patterns in real space as well (Fig.~\ref{figure3}c). Experimentally, we observe two LDOS maxima with a spacing not commensurate with the atomic lattice perpendicular to the island edges as shown in Fig.~\ref{figure3}d,e. Detailed comparison between the calculated and experimental LDOS suggests that the edges of the NiI$_2$ islands are magnetically more strongly coupled with the substrate (similar to the reported results on CrBr$_3$ and CrCl$_3$ \cite{Li2024ObservationHeterostructure, Cuperus2025Non-topologicalHeterostructures,Cuperus2025OneNbSe2}) (details shown in Supplementary Figs.~S3 and S7).

Additionally, we observe similar enhanced edge LDOS depends on all crystallographically equivivalent edges. The longer straight edges of the NiI$_2$ islands are at an angle of 60$^\circ$ and hence, are crystallographically equivalent (e.g. the ones shown in Fig.~\ref{figure3}d and e). All these edges feature the same double-stripe LDOS patterns. Some of the islands have shorter edge segments that form 120$^\circ$ angles with long edges. The stripe pattern is typically absent on these edges as illustrated in Supplementary Fig.~S2. While these two edge types have the same orientation w.r.t.~the NiI$_2$ lattice, they differ in whether the edge iodine atoms (assuming iodine terminated edges, which would be consistent with our growth conditions) lie above or below the plane of Ni atoms. This does conceivably affect the magnetic coupling with the substrate as shown by our experimental results. 

Theoretical LDOS maps in Fig.~\ref{figure3}c show that spin spiral ground state that breaks the three-fold symmetry of the system and should result in nonequivalent edge contrast between the crystallographically identical edges. This is in contrast with our experimental results that match those predicted for the ferromagnetic NiI$_2$ layer.

Finally, we have carried out more detailed spectroscopy experiments at sub-Kelvin temperatures that reveal several YSR states with varying intensities over the moiré period at the NiI$_2$ island edges as shown in the Fig.~\ref{figure3}f (See Supplementary Fig.~S3 for details). The superconducting gap hosts YSR states with different energies, including zero-bias peaks that display a modulation in their intensities over the moiré period. These observations are supported by the theoretical LDOS spectroscopies over the moiré period, shown in Fig.~\ref{figure3}g (See Supplementary Fig.~S7 and S8 for details). 

\section*{Discussion}
 
We have shown that NiI$_2$ grown on an NbSe$_2$ substrate undergoes a transition from its original helimagnetic to a ferromagnetic ground state.
Such a transition is driven by a substrate-induced
renormalization of the magnetic interactions,
stemming from the ferromagnetic RKKY
interaction of NbSe$_2$.
The presence of the ferromagnetic state is evident
by the absence of ferroelectric stripes in NiI$_2$.
Furthermore, we show that this ferromagnetic state is directly imprinted in the YSR states, both in the
bulk and edges of the NiI$_2$ islands.
The bulk YSR states
of the NiI$_2$ islands confirm the presence
of bulk magnetism.
Moreover, we observe enhanced YSR states at the island edges with a distinct spatial variation over the moir\'e period.
The symmetry of the edge modes allows us to further confirm
that the NiI$_2$ ground state is ferromagnetic, instead
of the original spin spiral state.
These results demonstrate the control of magnetic order in 2D magnet/superconductor heterostructures, and establish NiI$_2$
as a substrate-tunable frustrated magnet. Looking beyond the NiI$_2$/NbSe$_2$ system, our results show that YSR states in magnetic layers on superconductors are a sensitive tool that can reveal details of the magnet's ground state in a spin-averaged measurement.

\section*{Methods and Materials}

\subsection*{Sample preparation} 

NiI$_2$ is grown on NbSe$_2$ substrate using MBE in UHV conditions. NbSe$_2$ is cleaved in vacuum to get clean substrate, followed by annealing at 350\,$^{\circ}$C for 3 hours. During sample growth, the substrate temperature is kept at 120\,$^{\circ}$C. Since NiI$_2$ is not stable at the sublimation temperature of the powder, we used two separate sources for depositing nickel and iodine on the substrate \cite{Amini2024AtomicScalesub2/sub}. As metal source, we used nickel rod, and the nickel atoms are deposited onto the substrate  by using electron beam evaporator. We used NiI$_2$ powder with 99.999\% purity as I$_2$ source. To decompose Ni and I$_2$, the NiI$_2$ powder crucible was heated up to 450\,$^{\circ}$C. After the growth, we annealed the sample for another 10 minutes at growth temperature in iodine environment, to ensure the formation of large area islands. 

\subsection*{STM measurements}

After the preparation, the sample was inserted into the low-temperature STM connected to the same UHV system. The STM experiments were performed either at $T=5$~K (Createc LT-STM) or at $T=350$~mK (Unisoku USM1300 STM). STM images were taken in the constant-current mode. d$I$/d$V$ spectra were recorded by standard lock-in detection while sweeping the sample bias in an open feedback loop configuration, with a bias modulation amplitude specified for each measurement.

\section*{Data availability}
All the data supporting the findings are available from the corresponding authors upon request.

\section*{Acknowledgements}
This research made use of the Aalto Nanomicroscopy Center (Aalto NMC) facilities and was supported by the Research Council of Finland Academy Research Fellowships Nos.~347266, 368478, 371757, 369367, EU Horizon Europe Marie Skłodowska-Curie Action 101154353, ERC AdG GETREAL (no.~101142364), and ERC CoG ULTRATWISTROICS (no.~101170477). We acknowledge the financial support of the Finnish Ministry of Education and Culture through the Quantum Doctoral Education Pilot Program (QDOC VN/3137/2024-OKM-4), the Research Council of Finland through the Finnish Quantum Flagship project (358877, Aalto University), the Finnish Centre of Excellence in Quantum Materials QMAT (No. 374166), and the computational resources provided by the Aalto Science-IT project.

\section*{Contributions}

B.G.A, M.A., and P.L. initiated and conceptualized this project. B.G.A., Z.W., M.A., A.C. and R.D. performed the sample growth and carried out the STM/STS measurements. A.O.F., A.O. and J.L.L. performed the theoretical modelling. B.G.A. and R.D. performed the data analysis. B.G.A., A.O.F., J.L.L. and P.L. wrote the manuscript with feedback from all the authors.

\section*{Competing interests}
The authors declare no competing interests.

\bibliography{references,biblio}

\begin{thebibliography}{36}%
\makeatletter
\providecommand \@ifxundefined [1]{%
 \@ifx{#1\undefined}
}%
\providecommand \@ifnum [1]{%
 \ifnum #1\expandafter \@firstoftwo
 \else \expandafter \@secondoftwo
 \fi
}%
\providecommand \@ifx [1]{%
 \ifx #1\expandafter \@firstoftwo
 \else \expandafter \@secondoftwo
 \fi
}%
\providecommand \natexlab [1]{#1}%
\providecommand \enquote  [1]{``#1''}%
\providecommand \bibnamefont  [1]{#1}%
\providecommand \bibfnamefont [1]{#1}%
\providecommand \citenamefont [1]{#1}%
\providecommand \href@noop [0]{\@secondoftwo}%
\providecommand \href [0]{\begingroup \@sanitize@url \@href}%
\providecommand \@href[1]{\@@startlink{#1}\@@href}%
\providecommand \@@href[1]{\endgroup#1\@@endlink}%
\providecommand \@sanitize@url [0]{\catcode `\\12\catcode `\$12\catcode `\&12\catcode `\#12\catcode `\^12\catcode `\_12\catcode `\%12\relax}%
\providecommand \@@startlink[1]{}%
\providecommand \@@endlink[0]{}%
\providecommand \url  [0]{\begingroup\@sanitize@url \@url }%
\providecommand \@url [1]{\endgroup\@href {#1}{\urlprefix }}%
\providecommand \urlprefix  [0]{URL }%
\providecommand \Eprint [0]{\href }%
\providecommand \doibase [0]{http://dx.doi.org/}%
\providecommand \selectlanguage [0]{\@gobble}%
\providecommand \bibinfo  [0]{\@secondoftwo}%
\providecommand \bibfield  [0]{\@secondoftwo}%
\providecommand \translation [1]{[#1]}%
\providecommand \BibitemOpen [0]{}%
\providecommand \bibitemStop [0]{}%
\providecommand \bibitemNoStop [0]{.\EOS\space}%
\providecommand \EOS [0]{\spacefactor3000\relax}%
\providecommand \BibitemShut  [1]{\csname bibitem#1\endcsname}%
\let\auto@bib@innerbib\@empty
\bibitem [{\citenamefont {Gong}\ \emph {et~al.}(2017)\citenamefont {Gong}, \citenamefont {Li}, \citenamefont {Li}, \citenamefont {Ji}, \citenamefont {Stern}, \citenamefont {Xia}, \citenamefont {Cao}, \citenamefont {Bao}, \citenamefont {Wang}, \citenamefont {Wang}, \citenamefont {Qiu}, \citenamefont {Cava}, \citenamefont {Louie}, \citenamefont {Xia},\ and\ \citenamefont {Zhang}}]{Gong2017DiscoveryCrystals}%
  \BibitemOpen
  \bibfield  {author} {\bibinfo {author} {\bibfnamefont {Cheng}\ \bibnamefont {Gong}}, \bibinfo {author} {\bibfnamefont {Lin}\ \bibnamefont {Li}}, \bibinfo {author} {\bibfnamefont {Zhenglu}\ \bibnamefont {Li}}, \bibinfo {author} {\bibfnamefont {Huiwen}\ \bibnamefont {Ji}}, \bibinfo {author} {\bibfnamefont {Alex}\ \bibnamefont {Stern}}, \bibinfo {author} {\bibfnamefont {Yang}\ \bibnamefont {Xia}}, \bibinfo {author} {\bibfnamefont {Ting}\ \bibnamefont {Cao}}, \bibinfo {author} {\bibfnamefont {Wei}\ \bibnamefont {Bao}}, \bibinfo {author} {\bibfnamefont {Chenzhe}\ \bibnamefont {Wang}}, \bibinfo {author} {\bibfnamefont {Yuan}\ \bibnamefont {Wang}}, \bibinfo {author} {\bibfnamefont {Z.~Q.}\ \bibnamefont {Qiu}}, \bibinfo {author} {\bibfnamefont {R.~J.}\ \bibnamefont {Cava}}, \bibinfo {author} {\bibfnamefont {Steven~G.}\ \bibnamefont {Louie}}, \bibinfo {author} {\bibfnamefont {Jing}\ \bibnamefont {Xia}}, \ and\ \bibinfo {author} {\bibfnamefont {Xiang}\ \bibnamefont {Zhang}},\ }\bibfield  {title} {\enquote {\bibinfo
  {title} {{Discovery of intrinsic ferromagnetism in two-dimensional van der Waals crystals}},}\ }\href {\doibase 10.1038/nature22060} {\bibfield  {journal} {\bibinfo  {journal} {Nature}\ }\textbf {\bibinfo {volume} {546}},\ \bibinfo {pages} {265--269} (\bibinfo {year} {2017})}\BibitemShut {NoStop}%
\bibitem [{\citenamefont {Huang}\ \emph {et~al.}(2017)\citenamefont {Huang}, \citenamefont {Clark}, \citenamefont {Navarro-Moratalla}, \citenamefont {Klein}, \citenamefont {Cheng}, \citenamefont {Seyler}, \citenamefont {Zhong}, \citenamefont {Schmidgall}, \citenamefont {McGuire}, \citenamefont {Cobden}, \citenamefont {Yao}, \citenamefont {Xiao}, \citenamefont {Jarillo-Herrero},\ and\ \citenamefont {Xu}}]{Huang2017Layer-dependentLimit}%
  \BibitemOpen
  \bibfield  {author} {\bibinfo {author} {\bibfnamefont {Bevin}\ \bibnamefont {Huang}}, \bibinfo {author} {\bibfnamefont {Genevieve}\ \bibnamefont {Clark}}, \bibinfo {author} {\bibfnamefont {Efrén}\ \bibnamefont {Navarro-Moratalla}}, \bibinfo {author} {\bibfnamefont {Dahlia~R.}\ \bibnamefont {Klein}}, \bibinfo {author} {\bibfnamefont {Ran}\ \bibnamefont {Cheng}}, \bibinfo {author} {\bibfnamefont {Kyle~L.}\ \bibnamefont {Seyler}}, \bibinfo {author} {\bibfnamefont {Ding}\ \bibnamefont {Zhong}}, \bibinfo {author} {\bibfnamefont {Emma}\ \bibnamefont {Schmidgall}}, \bibinfo {author} {\bibfnamefont {Michael~A.}\ \bibnamefont {McGuire}}, \bibinfo {author} {\bibfnamefont {David~H.}\ \bibnamefont {Cobden}}, \bibinfo {author} {\bibfnamefont {Wang}\ \bibnamefont {Yao}}, \bibinfo {author} {\bibfnamefont {Di}~\bibnamefont {Xiao}}, \bibinfo {author} {\bibfnamefont {Pablo}\ \bibnamefont {Jarillo-Herrero}}, \ and\ \bibinfo {author} {\bibfnamefont {Xiaodong}\ \bibnamefont {Xu}},\ }\bibfield  {title} {\enquote {\bibinfo
  {title} {{Layer-dependent ferromagnetism in a van der Waals crystal down to the monolayer limit}},}\ }\href {\doibase 10.1038/nature22391} {\bibfield  {journal} {\bibinfo  {journal} {Nature}\ }\textbf {\bibinfo {volume} {546}},\ \bibinfo {pages} {270--273} (\bibinfo {year} {2017})}\BibitemShut {NoStop}%
\bibitem [{\citenamefont {Fei}\ \emph {et~al.}(2018)\citenamefont {Fei}, \citenamefont {Huang}, \citenamefont {Malinowski}, \citenamefont {Wang}, \citenamefont {Song}, \citenamefont {Sanchez}, \citenamefont {Yao}, \citenamefont {Xiao}, \citenamefont {Zhu}, \citenamefont {May}, \citenamefont {Wu}, \citenamefont {Cobden}, \citenamefont {Chu},\ and\ \citenamefont {Xu}}]{Fei2018Two-dimensionalFe3GeTe2}%
  \BibitemOpen
  \bibfield  {author} {\bibinfo {author} {\bibfnamefont {Zaiyao}\ \bibnamefont {Fei}}, \bibinfo {author} {\bibfnamefont {Bevin}\ \bibnamefont {Huang}}, \bibinfo {author} {\bibfnamefont {Paul}\ \bibnamefont {Malinowski}}, \bibinfo {author} {\bibfnamefont {Wenbo}\ \bibnamefont {Wang}}, \bibinfo {author} {\bibfnamefont {Tiancheng}\ \bibnamefont {Song}}, \bibinfo {author} {\bibfnamefont {Joshua}\ \bibnamefont {Sanchez}}, \bibinfo {author} {\bibfnamefont {Wang}\ \bibnamefont {Yao}}, \bibinfo {author} {\bibfnamefont {Di}~\bibnamefont {Xiao}}, \bibinfo {author} {\bibfnamefont {Xiaoyang}\ \bibnamefont {Zhu}}, \bibinfo {author} {\bibfnamefont {Andrew~F.}\ \bibnamefont {May}}, \bibinfo {author} {\bibfnamefont {Weida}\ \bibnamefont {Wu}}, \bibinfo {author} {\bibfnamefont {David~H.}\ \bibnamefont {Cobden}}, \bibinfo {author} {\bibfnamefont {Jiun-Haw}\ \bibnamefont {Chu}}, \ and\ \bibinfo {author} {\bibfnamefont {Xiaodong}\ \bibnamefont {Xu}},\ }\bibfield  {title} {\enquote {\bibinfo {title} {{Two-dimensional itinerant
  ferromagnetism in atomically thin Fe$_3$GeTe$_2$}},}\ }\href {\doibase 10.1038/s41563-018-0149-7} {\bibfield  {journal} {\bibinfo  {journal} {Nature Materials}\ }\textbf {\bibinfo {volume} {17}},\ \bibinfo {pages} {778--782} (\bibinfo {year} {2018})}\BibitemShut {NoStop}%
\bibitem [{\citenamefont {Bedoya-Pinto}\ \emph {et~al.}(2021)\citenamefont {Bedoya-Pinto}, \citenamefont {Ji}, \citenamefont {Pandeya}, \citenamefont {Gargiani}, \citenamefont {Valvidares}, \citenamefont {Sessi}, \citenamefont {Taylor}, \citenamefont {Radu}, \citenamefont {Chang},\ and\ \citenamefont {Parkin}}]{Bedoya-Pinto2021IntrinsicMonolayer}%
  \BibitemOpen
  \bibfield  {author} {\bibinfo {author} {\bibfnamefont {Amilcar}\ \bibnamefont {Bedoya-Pinto}}, \bibinfo {author} {\bibfnamefont {Jing-Rong}\ \bibnamefont {Ji}}, \bibinfo {author} {\bibfnamefont {Avanindra~K.}\ \bibnamefont {Pandeya}}, \bibinfo {author} {\bibfnamefont {Pierluigi}\ \bibnamefont {Gargiani}}, \bibinfo {author} {\bibfnamefont {Manuel}\ \bibnamefont {Valvidares}}, \bibinfo {author} {\bibfnamefont {Paolo}\ \bibnamefont {Sessi}}, \bibinfo {author} {\bibfnamefont {James~M.}\ \bibnamefont {Taylor}}, \bibinfo {author} {\bibfnamefont {Florin}\ \bibnamefont {Radu}}, \bibinfo {author} {\bibfnamefont {Kai}\ \bibnamefont {Chang}}, \ and\ \bibinfo {author} {\bibfnamefont {Stuart S.~P.}\ \bibnamefont {Parkin}},\ }\bibfield  {title} {\enquote {\bibinfo {title} {{Intrinsic 2D-XY ferromagnetism in a van der Waals monolayer}},}\ }\href {\doibase 10.1126/science.abd5146} {\bibfield  {journal} {\bibinfo  {journal} {Science}\ }\textbf {\bibinfo {volume} {374}},\ \bibinfo {pages} {616--620} (\bibinfo {year}
  {2021})}\BibitemShut {NoStop}%
\bibitem [{\citenamefont {Jiang}\ \emph {et~al.}(2018)\citenamefont {Jiang}, \citenamefont {Li}, \citenamefont {Wang}, \citenamefont {Mak},\ and\ \citenamefont {Shan}}]{Jiang2018ControllingDoping}%
  \BibitemOpen
  \bibfield  {author} {\bibinfo {author} {\bibfnamefont {Shengwei}\ \bibnamefont {Jiang}}, \bibinfo {author} {\bibfnamefont {Lizhong}\ \bibnamefont {Li}}, \bibinfo {author} {\bibfnamefont {Zefang}\ \bibnamefont {Wang}}, \bibinfo {author} {\bibfnamefont {Kin~Fai}\ \bibnamefont {Mak}}, \ and\ \bibinfo {author} {\bibfnamefont {Jie}\ \bibnamefont {Shan}},\ }\bibfield  {title} {\enquote {\bibinfo {title} {{Controlling magnetism in 2D CrI$_3$ by electrostatic doping}},}\ }\href {\doibase 10.1038/s41565-018-0135-x} {\bibfield  {journal} {\bibinfo  {journal} {Nature Nanotechnology}\ }\textbf {\bibinfo {volume} {13}},\ \bibinfo {pages} {549--553} (\bibinfo {year} {2018})}\BibitemShut {NoStop}%
\bibitem [{\citenamefont {Burch}\ \emph {et~al.}(2018)\citenamefont {Burch}, \citenamefont {Mandrus},\ and\ \citenamefont {Park}}]{Burch2018MagnetismMaterials}%
  \BibitemOpen
  \bibfield  {author} {\bibinfo {author} {\bibfnamefont {Kenneth~S.}\ \bibnamefont {Burch}}, \bibinfo {author} {\bibfnamefont {David}\ \bibnamefont {Mandrus}}, \ and\ \bibinfo {author} {\bibfnamefont {Je-Geun}\ \bibnamefont {Park}},\ }\bibfield  {title} {\enquote {\bibinfo {title} {{Magnetism in two-dimensional van der Waals materials}},}\ }\href {\doibase 10.1038/s41586-018-0631-z} {\bibfield  {journal} {\bibinfo  {journal} {Nature}\ }\textbf {\bibinfo {volume} {563}},\ \bibinfo {pages} {47--52} (\bibinfo {year} {2018})}\BibitemShut {NoStop}%
\bibitem [{\citenamefont {Gibertini}\ \emph {et~al.}(2019)\citenamefont {Gibertini}, \citenamefont {Koperski}, \citenamefont {Morpurgo},\ and\ \citenamefont {Novoselov}}]{Gibertini2019MagneticHeterostructures}%
  \BibitemOpen
  \bibfield  {author} {\bibinfo {author} {\bibfnamefont {M.}~\bibnamefont {Gibertini}}, \bibinfo {author} {\bibfnamefont {M.}~\bibnamefont {Koperski}}, \bibinfo {author} {\bibfnamefont {A.~F.}\ \bibnamefont {Morpurgo}}, \ and\ \bibinfo {author} {\bibfnamefont {K.~S.}\ \bibnamefont {Novoselov}},\ }\bibfield  {title} {\enquote {\bibinfo {title} {{Magnetic 2D materials and heterostructures}},}\ }\href {\doibase 10.1038/s41565-019-0438-6} {\bibfield  {journal} {\bibinfo  {journal} {Nature Nanotechnology}\ }\textbf {\bibinfo {volume} {14}},\ \bibinfo {pages} {408--419} (\bibinfo {year} {2019})}\BibitemShut {NoStop}%
\bibitem [{\citenamefont {Grubi{\v{s}}i{\'{c}}-{\v{C}}abo}\ \emph {et~al.}(2025)\citenamefont {Grubi{\v{s}}i{\'{c}}-{\v{C}}abo}, \citenamefont {Guimar{\~{a}}es}, \citenamefont {Afanasiev}, \citenamefont {Garcia~Aguilar}, \citenamefont {Aguilera}, \citenamefont {Ali}, \citenamefont {Bhattacharyya}, \citenamefont {Blanter}, \citenamefont {Bosma}, \citenamefont {Cheng}, \citenamefont {Dan}, \citenamefont {Dash}, \citenamefont {Medina~Due{\~{n}}as}, \citenamefont {Fernandez-Rossier}, \citenamefont {Gibertini}, \citenamefont {Grytsiuk}, \citenamefont {Houmes}, \citenamefont {Isaeva}, \citenamefont {Knekna}, \citenamefont {Kole}, \citenamefont {Kurdi}, \citenamefont {Lado}, \citenamefont {Ma{\~{n}}as-Valero}, \citenamefont {Lopes}, \citenamefont {Marian}, \citenamefont {Na}, \citenamefont {Pabst}, \citenamefont {Barquero~Pierantoni}, \citenamefont {Regout}, \citenamefont {Reho}, \citenamefont {R{\"{o}}sner}, \citenamefont {Sanz}, \citenamefont {van~der Sar}, \citenamefont {S{\l}awi{\'{n}}ska}, \citenamefont
  {Verstraete}, \citenamefont {Waseem}, \citenamefont {van~der Zant}, \citenamefont {Zanolli},\ and\ \citenamefont {Soriano}}]{Grubisic-Cabo2025RoadmapMaterials}%
  \BibitemOpen
  \bibfield  {author} {\bibinfo {author} {\bibfnamefont {Antonija}\ \bibnamefont {Grubi{\v{s}}i{\'{c}}-{\v{C}}abo}}, \bibinfo {author} {\bibfnamefont {Marcos~H.D.}\ \bibnamefont {Guimar{\~{a}}es}}, \bibinfo {author} {\bibfnamefont {Dmytro}\ \bibnamefont {Afanasiev}}, \bibinfo {author} {\bibfnamefont {Jose~H.}\ \bibnamefont {Garcia~Aguilar}}, \bibinfo {author} {\bibfnamefont {Irene}\ \bibnamefont {Aguilera}}, \bibinfo {author} {\bibfnamefont {Mazhar~N.}\ \bibnamefont {Ali}}, \bibinfo {author} {\bibfnamefont {Semonti}\ \bibnamefont {Bhattacharyya}}, \bibinfo {author} {\bibfnamefont {Yaroslav~M.}\ \bibnamefont {Blanter}}, \bibinfo {author} {\bibfnamefont {Rixt}\ \bibnamefont {Bosma}}, \bibinfo {author} {\bibfnamefont {Zhiyuan}\ \bibnamefont {Cheng}}, \bibinfo {author} {\bibfnamefont {Zhiying}\ \bibnamefont {Dan}}, \bibinfo {author} {\bibfnamefont {Saroj~P.}\ \bibnamefont {Dash}}, \bibinfo {author} {\bibfnamefont {Joaquín}\ \bibnamefont {Medina~Due{\~{n}}as}}, \bibinfo {author} {\bibfnamefont {Joaquín}\
  \bibnamefont {Fernandez-Rossier}}, \bibinfo {author} {\bibfnamefont {Marco}\ \bibnamefont {Gibertini}}, \bibinfo {author} {\bibfnamefont {Sergii}\ \bibnamefont {Grytsiuk}}, \bibinfo {author} {\bibfnamefont {Maurits~J.A.}\ \bibnamefont {Houmes}}, \bibinfo {author} {\bibfnamefont {Anna}\ \bibnamefont {Isaeva}}, \bibinfo {author} {\bibfnamefont {Chrystalla}\ \bibnamefont {Knekna}}, \bibinfo {author} {\bibfnamefont {Arnold~H.}\ \bibnamefont {Kole}}, \bibinfo {author} {\bibfnamefont {Samer}\ \bibnamefont {Kurdi}}, \bibinfo {author} {\bibfnamefont {Jose~L.}\ \bibnamefont {Lado}}, \bibinfo {author} {\bibfnamefont {Samuel}\ \bibnamefont {Ma{\~{n}}as-Valero}}, \bibinfo {author} {\bibfnamefont {J.~Marcelo~J.}\ \bibnamefont {Lopes}}, \bibinfo {author} {\bibfnamefont {Damiano}\ \bibnamefont {Marian}}, \bibinfo {author} {\bibfnamefont {Mengxing}\ \bibnamefont {Na}}, \bibinfo {author} {\bibfnamefont {Falk}\ \bibnamefont {Pabst}}, \bibinfo {author} {\bibfnamefont {Sergio}\ \bibnamefont {Barquero~Pierantoni}}, \bibinfo
  {author} {\bibfnamefont {Mexx}\ \bibnamefont {Regout}}, \bibinfo {author} {\bibfnamefont {Riccardo}\ \bibnamefont {Reho}}, \bibinfo {author} {\bibfnamefont {Malte}\ \bibnamefont {R{\"{o}}sner}}, \bibinfo {author} {\bibfnamefont {David}\ \bibnamefont {Sanz}}, \bibinfo {author} {\bibfnamefont {Toeno}\ \bibnamefont {van~der Sar}}, \bibinfo {author} {\bibfnamefont {Jagoda}\ \bibnamefont {S{\l}awi{\'{n}}ska}}, \bibinfo {author} {\bibfnamefont {Matthieu~J.}\ \bibnamefont {Verstraete}}, \bibinfo {author} {\bibfnamefont {Muhammad}\ \bibnamefont {Waseem}}, \bibinfo {author} {\bibfnamefont {Herre~S.J.}\ \bibnamefont {van~der Zant}}, \bibinfo {author} {\bibfnamefont {Zeila}\ \bibnamefont {Zanolli}}, \ and\ \bibinfo {author} {\bibfnamefont {David}\ \bibnamefont {Soriano}},\ }\bibfield  {title} {\enquote {\bibinfo {title} {{Roadmap on quantum magnetic materials}},}\ }\href {\doibase 10.1088/2053-1583/adbe89} {\bibfield  {journal} {\bibinfo  {journal} {2D Materials}\ }\textbf {\bibinfo {volume} {12}},\ \bibinfo {pages}
  {031501} (\bibinfo {year} {2025})}\BibitemShut {NoStop}%
\bibitem [{\citenamefont {Tokura}\ \emph {et~al.}(2014)\citenamefont {Tokura}, \citenamefont {Seki},\ and\ \citenamefont {Nagaosa}}]{Tokura2014MultiferroicsOrigin}%
  \BibitemOpen
  \bibfield  {author} {\bibinfo {author} {\bibfnamefont {Yoshinori}\ \bibnamefont {Tokura}}, \bibinfo {author} {\bibfnamefont {Shinichiro}\ \bibnamefont {Seki}}, \ and\ \bibinfo {author} {\bibfnamefont {Naoto}\ \bibnamefont {Nagaosa}},\ }\bibfield  {title} {\enquote {\bibinfo {title} {{Multiferroics of spin origin}},}\ }\href {\doibase 10.1088/0034-4885/77/7/076501} {\bibfield  {journal} {\bibinfo  {journal} {Reports on Progress in Physics}\ }\textbf {\bibinfo {volume} {77}},\ \bibinfo {pages} {076501} (\bibinfo {year} {2014})}\BibitemShut {NoStop}%
\bibitem [{\citenamefont {Ju}\ \emph {et~al.}(2021)\citenamefont {Ju}, \citenamefont {Lee}, \citenamefont {Kim}, \citenamefont {Choi}, \citenamefont {Roh}, \citenamefont {Son}, \citenamefont {Park}, \citenamefont {Kim}, \citenamefont {Jung}, \citenamefont {Kim}, \citenamefont {Kim}, \citenamefont {Park},\ and\ \citenamefont {Lee}}]{Ju2021Possiblesub2/sub}%
  \BibitemOpen
  \bibfield  {author} {\bibinfo {author} {\bibfnamefont {Hwiin}\ \bibnamefont {Ju}}, \bibinfo {author} {\bibfnamefont {Youjin}\ \bibnamefont {Lee}}, \bibinfo {author} {\bibfnamefont {Kwang-Tak}\ \bibnamefont {Kim}}, \bibinfo {author} {\bibfnamefont {In~Hyeok}\ \bibnamefont {Choi}}, \bibinfo {author} {\bibfnamefont {Chang~Jae}\ \bibnamefont {Roh}}, \bibinfo {author} {\bibfnamefont {Suhan}\ \bibnamefont {Son}}, \bibinfo {author} {\bibfnamefont {Pyeongjae}\ \bibnamefont {Park}}, \bibinfo {author} {\bibfnamefont {Jae~Ha}\ \bibnamefont {Kim}}, \bibinfo {author} {\bibfnamefont {Taek~Sun}\ \bibnamefont {Jung}}, \bibinfo {author} {\bibfnamefont {Jae~Hoon}\ \bibnamefont {Kim}}, \bibinfo {author} {\bibfnamefont {Kee~Hoon}\ \bibnamefont {Kim}}, \bibinfo {author} {\bibfnamefont {Je-Geun}\ \bibnamefont {Park}}, \ and\ \bibinfo {author} {\bibfnamefont {Jong~Seok}\ \bibnamefont {Lee}},\ }\bibfield  {title} {\enquote {\bibinfo {title} {{Possible Persistence of Multiferroic Order down to Bilayer Limit of van der Waals
  Material NiI$_2$}},}\ }\href {\doibase 10.1021/acs.nanolett.1c01095} {\bibfield  {journal} {\bibinfo  {journal} {Nano Letters}\ }\textbf {\bibinfo {volume} {21}},\ \bibinfo {pages} {5126--5132} (\bibinfo {year} {2021})}\BibitemShut {NoStop}%
\bibitem [{\citenamefont {Song}\ \emph {et~al.}(2022)\citenamefont {Song}, \citenamefont {Occhialini}, \citenamefont {Erge{\c{c}}en}, \citenamefont {Ilyas}, \citenamefont {Amoroso}, \citenamefont {Barone}, \citenamefont {Kapeghian}, \citenamefont {Watanabe}, \citenamefont {Taniguchi}, \citenamefont {Botana}, \citenamefont {Picozzi}, \citenamefont {Gedik},\ and\ \citenamefont {Comin}}]{Song2022EvidenceMultiferroic}%
  \BibitemOpen
  \bibfield  {author} {\bibinfo {author} {\bibfnamefont {Qian}\ \bibnamefont {Song}}, \bibinfo {author} {\bibfnamefont {Connor~A.}\ \bibnamefont {Occhialini}}, \bibinfo {author} {\bibfnamefont {Emre}\ \bibnamefont {Erge{\c{c}}en}}, \bibinfo {author} {\bibfnamefont {Batyr}\ \bibnamefont {Ilyas}}, \bibinfo {author} {\bibfnamefont {Danila}\ \bibnamefont {Amoroso}}, \bibinfo {author} {\bibfnamefont {Paolo}\ \bibnamefont {Barone}}, \bibinfo {author} {\bibfnamefont {Jesse}\ \bibnamefont {Kapeghian}}, \bibinfo {author} {\bibfnamefont {Kenji}\ \bibnamefont {Watanabe}}, \bibinfo {author} {\bibfnamefont {Takashi}\ \bibnamefont {Taniguchi}}, \bibinfo {author} {\bibfnamefont {Antia~S.}\ \bibnamefont {Botana}}, \bibinfo {author} {\bibfnamefont {Silvia}\ \bibnamefont {Picozzi}}, \bibinfo {author} {\bibfnamefont {Nuh}\ \bibnamefont {Gedik}}, \ and\ \bibinfo {author} {\bibfnamefont {Riccardo}\ \bibnamefont {Comin}},\ }\bibfield  {title} {\enquote {\bibinfo {title} {{Evidence for a single-layer van der Waals multiferroic}},}\
  }\href {\doibase 10.1038/s41586-021-04337-x} {\bibfield  {journal} {\bibinfo  {journal} {Nature}\ }\textbf {\bibinfo {volume} {602}},\ \bibinfo {pages} {601--605} (\bibinfo {year} {2022})}\BibitemShut {NoStop}%
\bibitem [{\citenamefont {Fumega}\ and\ \citenamefont {Lado}(2022)}]{Fumega2022}%
  \BibitemOpen
  \bibfield  {author} {\bibinfo {author} {\bibfnamefont {Adolfo~O.}\ \bibnamefont {Fumega}}\ and\ \bibinfo {author} {\bibfnamefont {Jose~L}\ \bibnamefont {Lado}},\ }\bibfield  {title} {\enquote {\bibinfo {title} {Microscopic origin of multiferroic order in monolayer {NiI}$_2$},}\ }\href {\doibase 10.1088/2053-1583/ac4e9d} {\bibfield  {journal} {\bibinfo  {journal} {2D Materials}\ }\textbf {\bibinfo {volume} {9}},\ \bibinfo {pages} {025010} (\bibinfo {year} {2022})}\BibitemShut {NoStop}%
\bibitem [{\citenamefont {Amini}\ \emph {et~al.}(2024)\citenamefont {Amini}, \citenamefont {Fumega}, \citenamefont {Gonz{\'{a}}lez‐Herrero}, \citenamefont {Vaňo}, \citenamefont {Kezilebieke}, \citenamefont {Lado},\ and\ \citenamefont {Liljeroth}}]{Amini2024AtomicScalesub2/sub}%
  \BibitemOpen
  \bibfield  {author} {\bibinfo {author} {\bibfnamefont {Mohammad}\ \bibnamefont {Amini}}, \bibinfo {author} {\bibfnamefont {Adolfo~O.}\ \bibnamefont {Fumega}}, \bibinfo {author} {\bibfnamefont {Héctor}\ \bibnamefont {Gonz{\'{a}}lez‐Herrero}}, \bibinfo {author} {\bibfnamefont {Viliam}\ \bibnamefont {Vaňo}}, \bibinfo {author} {\bibfnamefont {Shawulienu}\ \bibnamefont {Kezilebieke}}, \bibinfo {author} {\bibfnamefont {Jose~L.}\ \bibnamefont {Lado}}, \ and\ \bibinfo {author} {\bibfnamefont {Peter}\ \bibnamefont {Liljeroth}},\ }\bibfield  {title} {\enquote {\bibinfo {title} {{Atomic‐Scale Visualization of Multiferroicity in Monolayer NiI$_2$}},}\ }\href {\doibase 10.1002/adma.202311342} {\bibfield  {journal} {\bibinfo  {journal} {Advanced Materials}\ }\textbf {\bibinfo {volume} {36}},\ \bibinfo {pages} {2311342} (\bibinfo {year} {2024})}\BibitemShut {NoStop}%
\bibitem [{\citenamefont {Cahlík}\ \emph {et~al.}(2026)\citenamefont {Cahlík}, \citenamefont {Karjasilta}, \citenamefont {Mishra}, \citenamefont {Drost}, \citenamefont {Amini}, \citenamefont {Arshad}, \citenamefont {Arslan},\ and\ \citenamefont {Liljeroth}}]{2601.20713}%
  \BibitemOpen
  \bibfield  {author} {\bibinfo {author} {\bibfnamefont {Aleš}\ \bibnamefont {Cahlík}}, \bibinfo {author} {\bibfnamefont {Antti}\ \bibnamefont {Karjasilta}}, \bibinfo {author} {\bibfnamefont {Anshika}\ \bibnamefont {Mishra}}, \bibinfo {author} {\bibfnamefont {Robert}\ \bibnamefont {Drost}}, \bibinfo {author} {\bibfnamefont {Mohammad}\ \bibnamefont {Amini}}, \bibinfo {author} {\bibfnamefont {Javaria}\ \bibnamefont {Arshad}}, \bibinfo {author} {\bibfnamefont {B\"{u}şra~Gamze}\ \bibnamefont {Arslan}}, \ and\ \bibinfo {author} {\bibfnamefont {Peter}\ \bibnamefont {Liljeroth}},\ }\bibfield  {title} {\enquote {\bibinfo {title} {Probing type‐{II} multiferroicity in monolayer {NiBr}$_2$},}\ }\href {\doibase 10.1002/advs.77081} {\bibfield  {journal} {\bibinfo  {journal} {Advanced Science}\ } (\bibinfo {year} {2026}),\ 10.1002/advs.77081}\BibitemShut {NoStop}%
\bibitem [{\citenamefont {Miao}\ \emph {et~al.}(2025)\citenamefont {Miao}, \citenamefont {Liu}, \citenamefont {Zhang}, \citenamefont {Zhou}, \citenamefont {Wang}, \citenamefont {Wang}, \citenamefont {Ji},\ and\ \citenamefont {Fu}}]{Miao2025}%
  \BibitemOpen
  \bibfield  {author} {\bibinfo {author} {\bibfnamefont {Mao-Peng}\ \bibnamefont {Miao}}, \bibinfo {author} {\bibfnamefont {Nanshu}\ \bibnamefont {Liu}}, \bibinfo {author} {\bibfnamefont {Wen-Hao}\ \bibnamefont {Zhang}}, \bibinfo {author} {\bibfnamefont {Jian-Wang}\ \bibnamefont {Zhou}}, \bibinfo {author} {\bibfnamefont {Dao-Bo}\ \bibnamefont {Wang}}, \bibinfo {author} {\bibfnamefont {Cong}\ \bibnamefont {Wang}}, \bibinfo {author} {\bibfnamefont {Wei}\ \bibnamefont {Ji}}, \ and\ \bibinfo {author} {\bibfnamefont {Ying-Shuang}\ \bibnamefont {Fu}},\ }\bibfield  {title} {\enquote {\bibinfo {title} {Spin-resolved imaging of atomic-scale helimagnetism in mono- and bilayer {NiI}$_2$},}\ }\href {\doibase 10.1073/pnas.2422868122} {\bibfield  {journal} {\bibinfo  {journal} {Proc. Nat. Acad. Sci.}\ }\textbf {\bibinfo {volume} {122}},\ \bibinfo {pages} {e2422868122} (\bibinfo {year} {2025})}\BibitemShut {NoStop}%
\bibitem [{\citenamefont {Wang}\ \emph {et~al.}(2026)\citenamefont {Wang}, \citenamefont {Jiang}, \citenamefont {Pan}, \citenamefont {Wang}, \citenamefont {Wang}, \citenamefont {Tian}, \citenamefont {Li}, \citenamefont {Zhao}, \citenamefont {Zhang}, \citenamefont {Wang}, \citenamefont {Yang}, \citenamefont {Xiang}, \citenamefont {Xu}, \citenamefont {Feng},\ and\ \citenamefont {Zhang}}]{Wang2026}%
  \BibitemOpen
  \bibfield  {author} {\bibinfo {author} {\bibfnamefont {Haitao}\ \bibnamefont {Wang}}, \bibinfo {author} {\bibfnamefont {Tianxing}\ \bibnamefont {Jiang}}, \bibinfo {author} {\bibfnamefont {Weiyi}\ \bibnamefont {Pan}}, \bibinfo {author} {\bibfnamefont {Xu}~\bibnamefont {Wang}}, \bibinfo {author} {\bibfnamefont {Hongyu}\ \bibnamefont {Wang}}, \bibinfo {author} {\bibfnamefont {Junchao}\ \bibnamefont {Tian}}, \bibinfo {author} {\bibfnamefont {Lianchuang}\ \bibnamefont {Li}}, \bibinfo {author} {\bibfnamefont {Dongming}\ \bibnamefont {Zhao}}, \bibinfo {author} {\bibfnamefont {Qingle}\ \bibnamefont {Zhang}}, \bibinfo {author} {\bibfnamefont {Chenxi}\ \bibnamefont {Wang}}, \bibinfo {author} {\bibfnamefont {Ying}\ \bibnamefont {Yang}}, \bibinfo {author} {\bibfnamefont {Hongjun}\ \bibnamefont {Xiang}}, \bibinfo {author} {\bibfnamefont {Changsong}\ \bibnamefont {Xu}}, \bibinfo {author} {\bibfnamefont {Donglai}\ \bibnamefont {Feng}}, \ and\ \bibinfo {author} {\bibfnamefont {Tong}\ \bibnamefont {Zhang}},\ }\bibfield
  {title} {\enquote {\bibinfo {title} {Microscopic evidence of spin-driven multiferroicity and topological spin textures in monolayer {NiI}$_{2}$},}\ }\href {\doibase 10.1103/4hzc-bm2f} {\bibfield  {journal} {\bibinfo  {journal} {Phys. Rev. Lett.}\ }\textbf {\bibinfo {volume} {136}},\ \bibinfo {pages} {026402} (\bibinfo {year} {2026})}\BibitemShut {NoStop}%
\bibitem [{\citenamefont {Kezilebieke}\ \emph {et~al.}(2020)\citenamefont {Kezilebieke}, \citenamefont {Huda}, \citenamefont {Vaňo}, \citenamefont {Aapro}, \citenamefont {Ganguli}, \citenamefont {Silveira}, \citenamefont {G{\l}odzik}, \citenamefont {Foster}, \citenamefont {Ojanen},\ and\ \citenamefont {Liljeroth}}]{Kezilebieke2020TopologicalHeterostructure}%
  \BibitemOpen
  \bibfield  {author} {\bibinfo {author} {\bibfnamefont {Shawulienu}\ \bibnamefont {Kezilebieke}}, \bibinfo {author} {\bibfnamefont {Md~Nurul}\ \bibnamefont {Huda}}, \bibinfo {author} {\bibfnamefont {Viliam}\ \bibnamefont {Vaňo}}, \bibinfo {author} {\bibfnamefont {Markus}\ \bibnamefont {Aapro}}, \bibinfo {author} {\bibfnamefont {Somesh~C.}\ \bibnamefont {Ganguli}}, \bibinfo {author} {\bibfnamefont {Orlando~J.}\ \bibnamefont {Silveira}}, \bibinfo {author} {\bibfnamefont {Szczepan}\ \bibnamefont {G{\l}odzik}}, \bibinfo {author} {\bibfnamefont {Adam~S.}\ \bibnamefont {Foster}}, \bibinfo {author} {\bibfnamefont {Teemu}\ \bibnamefont {Ojanen}}, \ and\ \bibinfo {author} {\bibfnamefont {Peter}\ \bibnamefont {Liljeroth}},\ }\bibfield  {title} {\enquote {\bibinfo {title} {{Topological superconductivity in a van der Waals heterostructure}},}\ }\href {\doibase 10.1038/s41586-020-2989-y} {\bibfield  {journal} {\bibinfo  {journal} {Nature}\ }\textbf {\bibinfo {volume} {588}},\ \bibinfo {pages} {424--428} (\bibinfo {year}
  {2020})}\BibitemShut {NoStop}%
\bibitem [{\citenamefont {Vaňo}\ \emph {et~al.}(2021)\citenamefont {Vaňo}, \citenamefont {Amini}, \citenamefont {Ganguli}, \citenamefont {Chen}, \citenamefont {Lado}, \citenamefont {Kezilebieke},\ and\ \citenamefont {Liljeroth}}]{Vano2021ArtificialHeterostructure}%
  \BibitemOpen
  \bibfield  {author} {\bibinfo {author} {\bibfnamefont {Viliam}\ \bibnamefont {Vaňo}}, \bibinfo {author} {\bibfnamefont {Mohammad}\ \bibnamefont {Amini}}, \bibinfo {author} {\bibfnamefont {Somesh~C.}\ \bibnamefont {Ganguli}}, \bibinfo {author} {\bibfnamefont {Guangze}\ \bibnamefont {Chen}}, \bibinfo {author} {\bibfnamefont {Jose~L.}\ \bibnamefont {Lado}}, \bibinfo {author} {\bibfnamefont {Shawulienu}\ \bibnamefont {Kezilebieke}}, \ and\ \bibinfo {author} {\bibfnamefont {Peter}\ \bibnamefont {Liljeroth}},\ }\bibfield  {title} {\enquote {\bibinfo {title} {{Artificial heavy fermions in a van der Waals heterostructure}},}\ }\href {\doibase 10.1038/s41586-021-04021-0} {\bibfield  {journal} {\bibinfo  {journal} {Nature}\ }\textbf {\bibinfo {volume} {599}},\ \bibinfo {pages} {582--586} (\bibinfo {year} {2021})}\BibitemShut {NoStop}%
\bibitem [{\citenamefont {Wan}\ \emph {et~al.}(2023)\citenamefont {Wan}, \citenamefont {Harsh}, \citenamefont {Meninno}, \citenamefont {Dreher}, \citenamefont {Sajan}, \citenamefont {Guo}, \citenamefont {Errea}, \citenamefont {de~Juan},\ and\ \citenamefont {Ugeda}}]{Wan2023EvidenceLattice}%
  \BibitemOpen
  \bibfield  {author} {\bibinfo {author} {\bibfnamefont {Wen}\ \bibnamefont {Wan}}, \bibinfo {author} {\bibfnamefont {Rishav}\ \bibnamefont {Harsh}}, \bibinfo {author} {\bibfnamefont {Antonella}\ \bibnamefont {Meninno}}, \bibinfo {author} {\bibfnamefont {Paul}\ \bibnamefont {Dreher}}, \bibinfo {author} {\bibfnamefont {Sandra}\ \bibnamefont {Sajan}}, \bibinfo {author} {\bibfnamefont {Haojie}\ \bibnamefont {Guo}}, \bibinfo {author} {\bibfnamefont {Ion}\ \bibnamefont {Errea}}, \bibinfo {author} {\bibfnamefont {Fernando}\ \bibnamefont {de~Juan}}, \ and\ \bibinfo {author} {\bibfnamefont {Miguel~M.}\ \bibnamefont {Ugeda}},\ }\bibfield  {title} {\enquote {\bibinfo {title} {{Evidence for ground state coherence in a two-dimensional Kondo lattice}},}\ }\href {\doibase 10.1038/s41467-023-42803-4} {\bibfield  {journal} {\bibinfo  {journal} {Nature Communications}\ }\textbf {\bibinfo {volume} {14}},\ \bibinfo {pages} {7005} (\bibinfo {year} {2023})}\BibitemShut {NoStop}%
\bibitem [{\citenamefont {Zhong}\ \emph {et~al.}(2017)\citenamefont {Zhong}, \citenamefont {Seyler}, \citenamefont {Linpeng}, \citenamefont {Cheng}, \citenamefont {Sivadas}, \citenamefont {Huang}, \citenamefont {Schmidgall}, \citenamefont {Taniguchi}, \citenamefont {Watanabe}, \citenamefont {McGuire}, \citenamefont {Yao}, \citenamefont {Xiao}, \citenamefont {Fu},\ and\ \citenamefont {Xu}}]{Zhong2017VanValleytronics}%
  \BibitemOpen
  \bibfield  {author} {\bibinfo {author} {\bibfnamefont {Ding}\ \bibnamefont {Zhong}}, \bibinfo {author} {\bibfnamefont {Kyle~L.}\ \bibnamefont {Seyler}}, \bibinfo {author} {\bibfnamefont {Xiayu}\ \bibnamefont {Linpeng}}, \bibinfo {author} {\bibfnamefont {Ran}\ \bibnamefont {Cheng}}, \bibinfo {author} {\bibfnamefont {Nikhil}\ \bibnamefont {Sivadas}}, \bibinfo {author} {\bibfnamefont {Bevin}\ \bibnamefont {Huang}}, \bibinfo {author} {\bibfnamefont {Emma}\ \bibnamefont {Schmidgall}}, \bibinfo {author} {\bibfnamefont {Takashi}\ \bibnamefont {Taniguchi}}, \bibinfo {author} {\bibfnamefont {Kenji}\ \bibnamefont {Watanabe}}, \bibinfo {author} {\bibfnamefont {Michael~A.}\ \bibnamefont {McGuire}}, \bibinfo {author} {\bibfnamefont {Wang}\ \bibnamefont {Yao}}, \bibinfo {author} {\bibfnamefont {Di}~\bibnamefont {Xiao}}, \bibinfo {author} {\bibfnamefont {Kai-Mei~C.}\ \bibnamefont {Fu}}, \ and\ \bibinfo {author} {\bibfnamefont {Xiaodong}\ \bibnamefont {Xu}},\ }\bibfield  {title} {\enquote {\bibinfo {title} {{Van der Waals
  engineering of ferromagnetic semiconductor heterostructures for spin and valleytronics}},}\ }\href {\doibase 10.1126/sciadv.1603113} {\bibfield  {journal} {\bibinfo  {journal} {Science Advances}\ }\textbf {\bibinfo {volume} {3}},\ \bibinfo {pages} {e1603113} (\bibinfo {year} {2017})}\BibitemShut {NoStop}%
\bibitem [{\citenamefont {Seyler}\ \emph {et~al.}(2018)\citenamefont {Seyler}, \citenamefont {Zhong}, \citenamefont {Huang}, \citenamefont {Linpeng}, \citenamefont {Wilson}, \citenamefont {Taniguchi}, \citenamefont {Watanabe}, \citenamefont {Yao}, \citenamefont {Xiao}, \citenamefont {McGuire}, \citenamefont {Fu},\ and\ \citenamefont {Xu}}]{Seyler2018ValleyHeterostructures}%
  \BibitemOpen
  \bibfield  {author} {\bibinfo {author} {\bibfnamefont {Kyle~L.}\ \bibnamefont {Seyler}}, \bibinfo {author} {\bibfnamefont {Ding}\ \bibnamefont {Zhong}}, \bibinfo {author} {\bibfnamefont {Bevin}\ \bibnamefont {Huang}}, \bibinfo {author} {\bibfnamefont {Xiayu}\ \bibnamefont {Linpeng}}, \bibinfo {author} {\bibfnamefont {Nathan~P.}\ \bibnamefont {Wilson}}, \bibinfo {author} {\bibfnamefont {Takashi}\ \bibnamefont {Taniguchi}}, \bibinfo {author} {\bibfnamefont {Kenji}\ \bibnamefont {Watanabe}}, \bibinfo {author} {\bibfnamefont {Wang}\ \bibnamefont {Yao}}, \bibinfo {author} {\bibfnamefont {Di}~\bibnamefont {Xiao}}, \bibinfo {author} {\bibfnamefont {Michael~A.}\ \bibnamefont {McGuire}}, \bibinfo {author} {\bibfnamefont {Kai-Mei~C.}\ \bibnamefont {Fu}}, \ and\ \bibinfo {author} {\bibfnamefont {Xiaodong}\ \bibnamefont {Xu}},\ }\bibfield  {title} {\enquote {\bibinfo {title} {{Valley Manipulation by Optically Tuning the Magnetic Proximity Effect in WSe$_2$/CrI$_3$ Heterostructures}},}\ }\href {\doibase
  10.1021/acs.nanolett.8b01105} {\bibfield  {journal} {\bibinfo  {journal} {Nano Letters}\ }\textbf {\bibinfo {volume} {18}},\ \bibinfo {pages} {3823--3828} (\bibinfo {year} {2018})}\BibitemShut {NoStop}%
\bibitem [{\citenamefont {Song}\ \emph {et~al.}(2021)\citenamefont {Song}, \citenamefont {Sun}, \citenamefont {Anderson}, \citenamefont {Wang}, \citenamefont {Qian}, \citenamefont {Taniguchi}, \citenamefont {Watanabe}, \citenamefont {McGuire}, \citenamefont {St{\"{o}}hr}, \citenamefont {Xiao}, \citenamefont {Cao}, \citenamefont {Wrachtrup},\ and\ \citenamefont {Xu}}]{Song2021DirectMagnets}%
  \BibitemOpen
  \bibfield  {author} {\bibinfo {author} {\bibfnamefont {Tiancheng}\ \bibnamefont {Song}}, \bibinfo {author} {\bibfnamefont {Qi-Chao}\ \bibnamefont {Sun}}, \bibinfo {author} {\bibfnamefont {Eric}\ \bibnamefont {Anderson}}, \bibinfo {author} {\bibfnamefont {Chong}\ \bibnamefont {Wang}}, \bibinfo {author} {\bibfnamefont {Jimin}\ \bibnamefont {Qian}}, \bibinfo {author} {\bibfnamefont {Takashi}\ \bibnamefont {Taniguchi}}, \bibinfo {author} {\bibfnamefont {Kenji}\ \bibnamefont {Watanabe}}, \bibinfo {author} {\bibfnamefont {Michael~A.}\ \bibnamefont {McGuire}}, \bibinfo {author} {\bibfnamefont {Rainer}\ \bibnamefont {St{\"{o}}hr}}, \bibinfo {author} {\bibfnamefont {Di}~\bibnamefont {Xiao}}, \bibinfo {author} {\bibfnamefont {Ting}\ \bibnamefont {Cao}}, \bibinfo {author} {\bibfnamefont {Jörg}\ \bibnamefont {Wrachtrup}}, \ and\ \bibinfo {author} {\bibfnamefont {Xiaodong}\ \bibnamefont {Xu}},\ }\bibfield  {title} {\enquote {\bibinfo {title} {{Direct visualization of magnetic domains and moir{\'{e}} magnetism in
  twisted 2D magnets}},}\ }\href {\doibase 10.1126/science.abj7478} {\bibfield  {journal} {\bibinfo  {journal} {Science}\ }\textbf {\bibinfo {volume} {374}},\ \bibinfo {pages} {1140--1144} (\bibinfo {year} {2021})}\BibitemShut {NoStop}%
\bibitem [{\citenamefont {Chen}\ \emph {et~al.}(2023)\citenamefont {Chen}, \citenamefont {Wang}, \citenamefont {Li}, \citenamefont {Hao}, \citenamefont {Cai}, \citenamefont {Dai}, \citenamefont {Chen}, \citenamefont {Xing}, \citenamefont {Liu}, \citenamefont {Wang}, \citenamefont {Zhai}, \citenamefont {Zhou},\ and\ \citenamefont {Han}}]{Chen2023ManipulationHeterostructures}%
  \BibitemOpen
  \bibfield  {author} {\bibinfo {author} {\bibfnamefont {Xiaodie}\ \bibnamefont {Chen}}, \bibinfo {author} {\bibfnamefont {Haoyun}\ \bibnamefont {Wang}}, \bibinfo {author} {\bibfnamefont {Manshi}\ \bibnamefont {Li}}, \bibinfo {author} {\bibfnamefont {Qinghua}\ \bibnamefont {Hao}}, \bibinfo {author} {\bibfnamefont {Menghao}\ \bibnamefont {Cai}}, \bibinfo {author} {\bibfnamefont {Hongwei}\ \bibnamefont {Dai}}, \bibinfo {author} {\bibfnamefont {Hongjing}\ \bibnamefont {Chen}}, \bibinfo {author} {\bibfnamefont {Yuntong}\ \bibnamefont {Xing}}, \bibinfo {author} {\bibfnamefont {Jie}\ \bibnamefont {Liu}}, \bibinfo {author} {\bibfnamefont {Xia}\ \bibnamefont {Wang}}, \bibinfo {author} {\bibfnamefont {Tianyou}\ \bibnamefont {Zhai}}, \bibinfo {author} {\bibfnamefont {Xing}\ \bibnamefont {Zhou}}, \ and\ \bibinfo {author} {\bibfnamefont {Jun‐Bo}\ \bibnamefont {Han}},\ }\bibfield  {title} {\enquote {\bibinfo {title} {{Manipulation and Optical Detection of Artificial Topological Phenomena in 2D Van der Waals
  Fe$_5$GeTe$_2$/MnPS$_3$ Heterostructures}},}\ }\href {\doibase 10.1002/advs.202207617} {\bibfield  {journal} {\bibinfo  {journal} {Advanced Science}\ }\textbf {\bibinfo {volume} {10}},\ \bibinfo {pages} {2207617} (\bibinfo {year} {2023})}\BibitemShut {NoStop}%
\bibitem [{\citenamefont {Heinrich}\ \emph {et~al.}(2018)\citenamefont {Heinrich}, \citenamefont {Pascual},\ and\ \citenamefont {Franke}}]{Heinrich2018SingleSuperconductors}%
  \BibitemOpen
  \bibfield  {author} {\bibinfo {author} {\bibfnamefont {Benjamin~W.}\ \bibnamefont {Heinrich}}, \bibinfo {author} {\bibfnamefont {Jose~I.}\ \bibnamefont {Pascual}}, \ and\ \bibinfo {author} {\bibfnamefont {Katharina~J.}\ \bibnamefont {Franke}},\ }\bibfield  {title} {\enquote {\bibinfo {title} {{Single magnetic adsorbates on s-wave superconductors}},}\ }\href {\doibase 10.1016/j.progsurf.2018.01.001} {\bibfield  {journal} {\bibinfo  {journal} {Progress in Surface Science}\ }\textbf {\bibinfo {volume} {93}},\ \bibinfo {pages} {1--19} (\bibinfo {year} {2018})}\BibitemShut {NoStop}%
\bibitem [{\citenamefont {Lo~Conte}\ \emph {et~al.}(2024)\citenamefont {Lo~Conte}, \citenamefont {Wiebe}, \citenamefont {Rachel}, \citenamefont {Morr},\ and\ \citenamefont {Wiesendanger}}]{LoConte2024Magnet-superconductorSuperconductivity}%
  \BibitemOpen
  \bibfield  {author} {\bibinfo {author} {\bibfnamefont {Roberto}\ \bibnamefont {Lo~Conte}}, \bibinfo {author} {\bibfnamefont {Jens}\ \bibnamefont {Wiebe}}, \bibinfo {author} {\bibfnamefont {Stephan}\ \bibnamefont {Rachel}}, \bibinfo {author} {\bibfnamefont {Dirk~K.}\ \bibnamefont {Morr}}, \ and\ \bibinfo {author} {\bibfnamefont {Roland}\ \bibnamefont {Wiesendanger}},\ }\bibfield  {title} {\enquote {\bibinfo {title} {{Magnet-superconductor hybrid quantum systems: a materials platform for topological superconductivity}},}\ }\href {\doibase 10.1007/s40766-024-00060-1} {\bibfield  {journal} {\bibinfo  {journal} {La Rivista del Nuovo Cimento}\ }\textbf {\bibinfo {volume} {47}},\ \bibinfo {pages} {453--554} (\bibinfo {year} {2024})}\BibitemShut {NoStop}%
\bibitem [{\citenamefont {Li}\ \emph {et~al.}(2024)\citenamefont {Li}, \citenamefont {Zhang}, \citenamefont {Zhu}, \citenamefont {Chen}, \citenamefont {Yi}, \citenamefont {Shen}, \citenamefont {Hou}, \citenamefont {Yan}, \citenamefont {Yao}, \citenamefont {Guo},\ and\ \citenamefont {Zhong}}]{Li2024ObservationHeterostructure}%
  \BibitemOpen
  \bibfield  {author} {\bibinfo {author} {\bibfnamefont {Peigen}\ \bibnamefont {Li}}, \bibinfo {author} {\bibfnamefont {Jihai}\ \bibnamefont {Zhang}}, \bibinfo {author} {\bibfnamefont {Di}~\bibnamefont {Zhu}}, \bibinfo {author} {\bibfnamefont {Cui-Qun}\ \bibnamefont {Chen}}, \bibinfo {author} {\bibfnamefont {Enkui}\ \bibnamefont {Yi}}, \bibinfo {author} {\bibfnamefont {Bing}\ \bibnamefont {Shen}}, \bibinfo {author} {\bibfnamefont {Yusheng}\ \bibnamefont {Hou}}, \bibinfo {author} {\bibfnamefont {Zhongbo}\ \bibnamefont {Yan}}, \bibinfo {author} {\bibfnamefont {Dao-Xin}\ \bibnamefont {Yao}}, \bibinfo {author} {\bibfnamefont {Donghui}\ \bibnamefont {Guo}}, \ and\ \bibinfo {author} {\bibfnamefont {Dingyong}\ \bibnamefont {Zhong}},\ }\bibfield  {title} {\enquote {\bibinfo {title} {{Observation of In-Gap States in a Two-Dimensional CrI$_2$/NbSe$_2$ Heterostructure}},}\ }\href {\doibase 10.1021/acs.nanolett.4c01848} {\bibfield  {journal} {\bibinfo  {journal} {Nano Letters}\ }\textbf {\bibinfo {volume} {24}},\
  \bibinfo {pages} {9468--9476} (\bibinfo {year} {2024})}\BibitemShut {NoStop}%
\bibitem [{\citenamefont {Cuperus}\ \emph {et~al.}(2025{\natexlab{a}})\citenamefont {Cuperus}, \citenamefont {Kole}, \citenamefont {Botello-M{\'{e}}ndez}, \citenamefont {Zanolli}, \citenamefont {Vanmaekelbergh},\ and\ \citenamefont {Swart}}]{Cuperus2025OneNbSe2}%
  \BibitemOpen
  \bibfield  {author} {\bibinfo {author} {\bibfnamefont {Jan~P.}\ \bibnamefont {Cuperus}}, \bibinfo {author} {\bibfnamefont {Arnold~H.}\ \bibnamefont {Kole}}, \bibinfo {author} {\bibfnamefont {Andrés~R.}\ \bibnamefont {Botello-M{\'{e}}ndez}}, \bibinfo {author} {\bibfnamefont {Zeila}\ \bibnamefont {Zanolli}}, \bibinfo {author} {\bibfnamefont {Daniel}\ \bibnamefont {Vanmaekelbergh}}, \ and\ \bibinfo {author} {\bibfnamefont {Ingmar}\ \bibnamefont {Swart}},\ }\bibfield  {title} {\enquote {\bibinfo {title} {{One dimensional edge localized YSR states in CrCl$_3$ on NbSe$_2$}},}\ }\href {\doibase 10.1038/s41535-025-00759-2} {\bibfield  {journal} {\bibinfo  {journal} {npj Quantum Materials}\ }\textbf {\bibinfo {volume} {10}},\ \bibinfo {pages} {51} (\bibinfo {year} {2025}{\natexlab{a}})}\BibitemShut {NoStop}%
\bibitem [{\citenamefont {Cuperus}\ \emph {et~al.}(2025{\natexlab{b}})\citenamefont {Cuperus}, \citenamefont {Vanmaekelbergh},\ and\ \citenamefont {Swart}}]{Cuperus2025Non-topologicalHeterostructures}%
  \BibitemOpen
  \bibfield  {author} {\bibinfo {author} {\bibfnamefont {Jan}\ \bibnamefont {Cuperus}}, \bibinfo {author} {\bibfnamefont {Daniël}\ \bibnamefont {Vanmaekelbergh}}, \ and\ \bibinfo {author} {\bibfnamefont {Ingmar}\ \bibnamefont {Swart}},\ }\bibfield  {title} {\enquote {\bibinfo {title} {{Non-topological edge-localized Yu-Shiba-Rusinov states in CrBr$_3$/NbSe$_2$ heterostructures}},}\ }\href {\doibase 10.21468/SciPostPhys.18.4.141} {\bibfield  {journal} {\bibinfo  {journal} {SciPost Physics}\ }\textbf {\bibinfo {volume} {18}},\ \bibinfo {pages} {141} (\bibinfo {year} {2025}{\natexlab{b}})}\BibitemShut {NoStop}%
\bibitem [{\citenamefont {Soldini}\ \emph {et~al.}(2023)\citenamefont {Soldini}, \citenamefont {K{\"{u}}ster}, \citenamefont {Wagner}, \citenamefont {Das}, \citenamefont {Aldarawsheh}, \citenamefont {Thomale}, \citenamefont {Lounis}, \citenamefont {Parkin}, \citenamefont {Sessi},\ and\ \citenamefont {Neupert}}]{Soldini2023Two-dimensionalSuperconductivity}%
  \BibitemOpen
  \bibfield  {author} {\bibinfo {author} {\bibfnamefont {Martina~O.}\ \bibnamefont {Soldini}}, \bibinfo {author} {\bibfnamefont {Felix}\ \bibnamefont {K{\"{u}}ster}}, \bibinfo {author} {\bibfnamefont {Glenn}\ \bibnamefont {Wagner}}, \bibinfo {author} {\bibfnamefont {Souvik}\ \bibnamefont {Das}}, \bibinfo {author} {\bibfnamefont {Amal}\ \bibnamefont {Aldarawsheh}}, \bibinfo {author} {\bibfnamefont {Ronny}\ \bibnamefont {Thomale}}, \bibinfo {author} {\bibfnamefont {Samir}\ \bibnamefont {Lounis}}, \bibinfo {author} {\bibfnamefont {Stuart S.~P.}\ \bibnamefont {Parkin}}, \bibinfo {author} {\bibfnamefont {Paolo}\ \bibnamefont {Sessi}}, \ and\ \bibinfo {author} {\bibfnamefont {Titus}\ \bibnamefont {Neupert}},\ }\bibfield  {title} {\enquote {\bibinfo {title} {{Two-dimensional Shiba lattices as a possible platform for crystalline topological superconductivity}},}\ }\href {\doibase 10.1038/s41567-023-02104-5} {\bibfield  {journal} {\bibinfo  {journal} {Nature Physics}\ }\textbf {\bibinfo {volume} {19}},\ \bibinfo
  {pages} {1848--1854} (\bibinfo {year} {2023})}\BibitemShut {NoStop}%
\bibitem [{\citenamefont {Wang}\ \emph {et~al.}(2024)\citenamefont {Wang}, \citenamefont {Zhao}, \citenamefont {Yao}, \citenamefont {Liu}, \citenamefont {Cheng}, \citenamefont {Zhang}, \citenamefont {Feng}, \citenamefont {Ma}, \citenamefont {Zhao}, \citenamefont {Sun}, \citenamefont {Wu},\ and\ \citenamefont {Chen}}]{Wang2024Orientation-selectiveNiI2}%
  \BibitemOpen
  \bibfield  {author} {\bibinfo {author} {\bibfnamefont {Yu}~\bibnamefont {Wang}}, \bibinfo {author} {\bibfnamefont {Xinlei}\ \bibnamefont {Zhao}}, \bibinfo {author} {\bibfnamefont {Li}~\bibnamefont {Yao}}, \bibinfo {author} {\bibfnamefont {Huiru}\ \bibnamefont {Liu}}, \bibinfo {author} {\bibfnamefont {Peng}\ \bibnamefont {Cheng}}, \bibinfo {author} {\bibfnamefont {Yiqi}\ \bibnamefont {Zhang}}, \bibinfo {author} {\bibfnamefont {Baojie}\ \bibnamefont {Feng}}, \bibinfo {author} {\bibfnamefont {Fengjie}\ \bibnamefont {Ma}}, \bibinfo {author} {\bibfnamefont {Jin}\ \bibnamefont {Zhao}}, \bibinfo {author} {\bibfnamefont {Jiatao}\ \bibnamefont {Sun}}, \bibinfo {author} {\bibfnamefont {Kehui}\ \bibnamefont {Wu}}, \ and\ \bibinfo {author} {\bibfnamefont {Lan}\ \bibnamefont {Chen}},\ }\bibfield  {title} {\enquote {\bibinfo {title} {{Orientation-selective spin-polarized edge states in monolayer NiI$_2$}},}\ }\href {\doibase 10.1038/s41467-024-55372-x} {\bibfield  {journal} {\bibinfo  {journal} {Nature Communications}\
  }\textbf {\bibinfo {volume} {15}},\ \bibinfo {pages} {10916} (\bibinfo {year} {2024})}\BibitemShut {NoStop}%
\bibitem [{\citenamefont {Wickramaratne}\ \emph {et~al.}(2020)\citenamefont {Wickramaratne}, \citenamefont {Khmelevskyi}, \citenamefont {Agterberg},\ and\ \citenamefont {Mazin}}]{PhysRevX.10.041003}%
  \BibitemOpen
  \bibfield  {author} {\bibinfo {author} {\bibfnamefont {Darshana}\ \bibnamefont {Wickramaratne}}, \bibinfo {author} {\bibfnamefont {Sergii}\ \bibnamefont {Khmelevskyi}}, \bibinfo {author} {\bibfnamefont {Daniel~F.}\ \bibnamefont {Agterberg}}, \ and\ \bibinfo {author} {\bibfnamefont {I.~I.}\ \bibnamefont {Mazin}},\ }\bibfield  {title} {\enquote {\bibinfo {title} {Ising superconductivity and magnetism in {NbSe}$_2$},}\ }\href {\doibase 10.1103/PhysRevX.10.041003} {\bibfield  {journal} {\bibinfo  {journal} {Phys. Rev. X}\ }\textbf {\bibinfo {volume} {10}},\ \bibinfo {pages} {041003} (\bibinfo {year} {2020})}\BibitemShut {NoStop}%
\bibitem [{\citenamefont {Schulz}\ \emph {et~al.}(2014)\citenamefont {Schulz}, \citenamefont {Drost}, \citenamefont {H{\"{a}}m{\"{a}}l{\"{a}}inen}, \citenamefont {Demonchaux}, \citenamefont {Seitsonen},\ and\ \citenamefont {Liljeroth}}]{Schulz2014EpitaxialTemplate}%
  \BibitemOpen
  \bibfield  {author} {\bibinfo {author} {\bibfnamefont {Fabian}\ \bibnamefont {Schulz}}, \bibinfo {author} {\bibfnamefont {Robert}\ \bibnamefont {Drost}}, \bibinfo {author} {\bibfnamefont {Sampsa~K.}\ \bibnamefont {H{\"{a}}m{\"{a}}l{\"{a}}inen}}, \bibinfo {author} {\bibfnamefont {Thomas}\ \bibnamefont {Demonchaux}}, \bibinfo {author} {\bibfnamefont {Ari~P.}\ \bibnamefont {Seitsonen}}, \ and\ \bibinfo {author} {\bibfnamefont {Peter}\ \bibnamefont {Liljeroth}},\ }\bibfield  {title} {\enquote {\bibinfo {title} {{Epitaxial hexagonal boron nitride on Ir(111): A work function template}},}\ }\href {\doibase 10.1103/PhysRevB.89.235429} {\bibfield  {journal} {\bibinfo  {journal} {Physical Review B}\ }\textbf {\bibinfo {volume} {89}},\ \bibinfo {pages} {235429} (\bibinfo {year} {2014})}\BibitemShut {NoStop}%
\bibitem [{\citenamefont {Joshi}\ \emph {et~al.}(2012)\citenamefont {Joshi}, \citenamefont {Ecija}, \citenamefont {Koitz}, \citenamefont {Iannuzzi}, \citenamefont {Seitsonen}, \citenamefont {Hutter}, \citenamefont {Sachdev}, \citenamefont {Vijayaraghavan}, \citenamefont {Bischoff}, \citenamefont {Seufert}, \citenamefont {Barth},\ and\ \citenamefont {Auw{\"{a}}rter}}]{Joshi2012BoronMonolayer}%
  \BibitemOpen
  \bibfield  {author} {\bibinfo {author} {\bibfnamefont {Sushobhan}\ \bibnamefont {Joshi}}, \bibinfo {author} {\bibfnamefont {David}\ \bibnamefont {Ecija}}, \bibinfo {author} {\bibfnamefont {Ralph}\ \bibnamefont {Koitz}}, \bibinfo {author} {\bibfnamefont {Marcella}\ \bibnamefont {Iannuzzi}}, \bibinfo {author} {\bibfnamefont {Ari~P.}\ \bibnamefont {Seitsonen}}, \bibinfo {author} {\bibfnamefont {Jürg}\ \bibnamefont {Hutter}}, \bibinfo {author} {\bibfnamefont {Hermann}\ \bibnamefont {Sachdev}}, \bibinfo {author} {\bibfnamefont {Saranyan}\ \bibnamefont {Vijayaraghavan}}, \bibinfo {author} {\bibfnamefont {Felix}\ \bibnamefont {Bischoff}}, \bibinfo {author} {\bibfnamefont {Knud}\ \bibnamefont {Seufert}}, \bibinfo {author} {\bibfnamefont {Johannes~V.}\ \bibnamefont {Barth}}, \ and\ \bibinfo {author} {\bibfnamefont {Willi}\ \bibnamefont {Auw{\"{a}}rter}},\ }\bibfield  {title} {\enquote {\bibinfo {title} {{Boron Nitride on Cu(111): An Electronically Corrugated Monolayer}},}\ }\href {\doibase 10.1021/nl303170m}
  {\bibfield  {journal} {\bibinfo  {journal} {Nano Letters}\ }\textbf {\bibinfo {volume} {12}},\ \bibinfo {pages} {5821--5828} (\bibinfo {year} {2012})}\BibitemShut {NoStop}%
\bibitem [{\citenamefont {J.~R.~Costa}\ \emph {et~al.}(2025)\citenamefont {J.~R.~Costa}, \citenamefont {Arribas}, \citenamefont {G.~L.~Brito}, \citenamefont {Cheng}, \citenamefont {Bradford}, \citenamefont {Thompson}, \citenamefont {Saywell}, \citenamefont {Mellor}, \citenamefont {Beton}, \citenamefont {Novikov}, \citenamefont {Plo}, \citenamefont {Gil}, \citenamefont {Cassabois}, \citenamefont {Zagonel}, \citenamefont {Kuhnke}, \citenamefont {Kern},\ and\ \citenamefont {Ros{\l}awska}}]{J.R.Costa2025NanoscaleGraphite}%
  \BibitemOpen
  \bibfield  {author} {\bibinfo {author} {\bibfnamefont {Fábio}\ \bibnamefont {J.~R.~Costa}}, \bibinfo {author} {\bibfnamefont {Daniel}\ \bibnamefont {Arribas}}, \bibinfo {author} {\bibfnamefont {Thiago}\ \bibnamefont {G.~L.~Brito}}, \bibinfo {author} {\bibfnamefont {Tin~S.}\ \bibnamefont {Cheng}}, \bibinfo {author} {\bibfnamefont {Jonathan}\ \bibnamefont {Bradford}}, \bibinfo {author} {\bibfnamefont {Amelia}\ \bibnamefont {Thompson}}, \bibinfo {author} {\bibfnamefont {Alex}\ \bibnamefont {Saywell}}, \bibinfo {author} {\bibfnamefont {Christopher~J.}\ \bibnamefont {Mellor}}, \bibinfo {author} {\bibfnamefont {Peter~H.}\ \bibnamefont {Beton}}, \bibinfo {author} {\bibfnamefont {Sergei~V.}\ \bibnamefont {Novikov}}, \bibinfo {author} {\bibfnamefont {Juliette}\ \bibnamefont {Plo}}, \bibinfo {author} {\bibfnamefont {Bernard}\ \bibnamefont {Gil}}, \bibinfo {author} {\bibfnamefont {Guillaume}\ \bibnamefont {Cassabois}}, \bibinfo {author} {\bibfnamefont {Luiz~Fernando}\ \bibnamefont {Zagonel}}, \bibinfo {author}
  {\bibfnamefont {Klaus}\ \bibnamefont {Kuhnke}}, \bibinfo {author} {\bibfnamefont {Klaus}\ \bibnamefont {Kern}}, \ and\ \bibinfo {author} {\bibfnamefont {Anna}\ \bibnamefont {Ros{\l}awska}},\ }\bibfield  {title} {\enquote {\bibinfo {title} {{Nanoscale Band Gap Modulation and Dual Moir{\'{e}} Superlattices of Hexagonal Boron Nitride Weakly Coupled to Graphite}},}\ }\href {\doibase 10.1021/acsnano.5c09374} {\bibfield  {journal} {\bibinfo  {journal} {ACS Nano}\ }\textbf {\bibinfo {volume} {19}},\ \bibinfo {pages} {35528--35538} (\bibinfo {year} {2025})}\BibitemShut {NoStop}%
\bibitem [{\citenamefont {Kezilebieke}\ \emph {et~al.}(2022)\citenamefont {Kezilebieke}, \citenamefont {Vaňo}, \citenamefont {Huda}, \citenamefont {Aapro}, \citenamefont {Ganguli}, \citenamefont {Liljeroth},\ and\ \citenamefont {Lado}}]{Kezilebieke2022}%
  \BibitemOpen
  \bibfield  {author} {\bibinfo {author} {\bibfnamefont {Shawulienu}\ \bibnamefont {Kezilebieke}}, \bibinfo {author} {\bibfnamefont {Viliam}\ \bibnamefont {Vaňo}}, \bibinfo {author} {\bibfnamefont {Md~N.}\ \bibnamefont {Huda}}, \bibinfo {author} {\bibfnamefont {Markus}\ \bibnamefont {Aapro}}, \bibinfo {author} {\bibfnamefont {Somesh~C.}\ \bibnamefont {Ganguli}}, \bibinfo {author} {\bibfnamefont {Peter}\ \bibnamefont {Liljeroth}}, \ and\ \bibinfo {author} {\bibfnamefont {Jose~L.}\ \bibnamefont {Lado}},\ }\bibfield  {title} {\enquote {\bibinfo {title} {Moiré-enabled topological superconductivity},}\ }\href {\doibase 10.1021/acs.nanolett.1c03856} {\bibfield  {journal} {\bibinfo  {journal} {Nano Letters}\ }\textbf {\bibinfo {volume} {22}},\ \bibinfo {pages} {328–333} (\bibinfo {year} {2022})}\BibitemShut {NoStop}%
\bibitem [{\citenamefont {Lado}(2026)}]{pyqula}%
  \BibitemOpen
  \bibfield  {author} {\bibinfo {author} {\bibfnamefont {Jose~L.}\ \bibnamefont {Lado}},\ }\href {https://github.com/joselado/pyqula} {\enquote {\bibinfo {title} {https://github.com/joselado/pyqula},}\ } (\bibinfo {year} {2026})\BibitemShut {NoStop}%
\end{thebibliography}%

\end{document}